\documentclass[11pt]{article}

\include{setup}

\begin{document}

\thispagestyle{empty}
\def\thefootnote{\fnsymbol{footnote}}
\setcounter{footnote}{1}
\null
\draftdate\hfill FR-PHENO-2011-004\\
\strut\hfill SFB/CPP-11-07\\
\strut\hfill TTK-11-05\\
\strut\hfill TTP11-04\\
\strut\hfill LPN11-37\\
\vskip 0cm
\vfill
\begin{center}
  {\Large \boldmath{\bf Electroweak corrections to dilepton + jet
      production \\  at hadron colliders}
\par} \vskip 2.5em
{\large
{\sc Ansgar Denner$^{1}$, Stefan Dittmaier$^{2}$, 
     Tobias Kasprzik$^{3}$, Alexander M\"uck$^{4}$
}\\[2ex]
{\normalsize \it $^1$Universit\"at W\"urzburg, 
Institut f\"ur Theoretische Physik und Astrophysik,\\ 
D-97074 W\"urzburg, Germany}
\\[1ex]
{\normalsize \it 
$^2$Albert-Ludwigs-Universit\"at Freiburg, 
Physikalisches Institut, \\
D-79104 Freiburg, Germany
}\\[1ex]
{\normalsize \it 
$^3$Karlsruhe Institute of Technology (KIT), 
Institut f\"ur Theoretische Teilchenphysik,\\
D-76128 Karlsruhe, Germany
}\\[1ex]
{\normalsize \it 
$^4$RWTH Aachen University, 
Institut f\"ur Theoretische Teilchenphysik und Kosmologie,\\
D-52056 Aachen, Germany
}\\[2ex]
}
\par \vskip 1em
\end{center}\par
\vskip .0cm \vfill {\bf Abstract:} \par
The first calculation of the next-to-leading-order 
electroweak corrections to Z-boson + jet hadroproduction
including leptonic Z-boson decays is presented, i.e.\ to the production of a
charged lepton--anti-lepton final state in association with one hard
jet at the LHC and the Tevatron. The \PZ-boson resonance is treated
consistently using the complex-mass scheme, and all off-shell effects
as well as the contributions of the intermediate photon are taken into
account. The corresponding next-to-leading-order QCD corrections have
also been recalculated. The full calculation is implemented in a
flexible Monte Carlo code. Numerical results for cross sections and
distributions of this Standard Model benchmark process are presented
for the Tevatron and the LHC.

\par
\vskip 1cm
\noindent
July 2011
\par
\null
\setcounter{page}{0}
\clearpage
\def\thefootnote{\arabic{footnote}}
\setcounter{footnote}{0}

\section{Introduction}

The Drell--Yan process is a cornerstone of electroweak (EW) physics
at hadron colliders like the Tevatron and the LHC. The
production of \PW\ and \PZ\ bosons (or off-shell photons) with subsequent
leptonic decays has both clean signatures and
large cross sections. It can potentially be used to measure the luminosity 
of the collider, to constrain the PDFs, or to calibrate the detector.
On the theoretical side, it is a perfect testing ground for our
understanding of hadron-collider physics.
Combining experimental accuracy and theoretical predictivity allows
for a number of precision measurements in spite of the hadron-collider
environment. (See e.g.\ Refs.~\cite{Haywood:1999qg,Gerber:2007xk} and
references therein.)

At hadron colliders, the EW gauge bosons are almost always
produced together with additional QCD radiation. In this work, we focus
on the neutral-current Drell--Yan process in which the dilepton pair is
produced in association with a hard, visible jet, i.e.\ 
\beq
\Pp\Pp/\Pp\bar\Pp \to \PZ/\gamma^* + \mathrm{jet} \to \Pl^+\Pl^- +
\mathrm{jet} +\X .  
\eeq 
The production cross section of this process,
which is widely dominated by resonant \PZ\ bosons, is large, and the
final state of the hard-scattering process is completely
reconstructable.  When the transverse momentum of the jet is large,
the dilepton pair will be boosted, and the process provides a source
for high-energy lepton pairs with opposite charge (opposite-sign
lepton pairs).  The invariant-mass distribution of the leptons is
dominated by the \PZ-boson resonance, and a good theoretical
understanding of this resonance also for boosted Z~bosons is a
cornerstone of an efficient detector calibration.
Providing high-energy lepton pairs and jet(s), $\PZ+\mathrm{jet(s)}$
production is not only a Standard Model (SM) candle process but
also an important background for new-physics searches. Moreover, the
process offers the possibility for precision tests of jet
dynamics in QCD.

The importance of Drell--Yan processes is also reflected in the
effort to make the theoretical predictions as precise as possible. The
differential cross section for \PW/\PZ\ production is known at
next-to-next-to-leading order (NNLO) accuracy (i.e.\ at two loops)
with respect to QCD corrections~\cite{vanNeerven:1991gh} and even up
to N$^3$LO in the soft-plus-virtual approximation~\cite{Moch:2005ky}.
The next-to-leading-order (NLO) QCD corrections have been matched with
parton showers \cite{Frixione:2006gn} and combined with a summation of
soft-gluon radiation (see e.g.\ \citere{Arnold:1990yk}), which is
necessary to predict the transverse-momentum distribution of the EW
bosons for small $p_{\rT}$.  In view of QCD only, the neutral- and
charged-current Drell--Yan processes are very similar. However,
concerning EW corrections, the production of \PW\ and \PZ\ bosons show
considerable differences and have been addressed separately. The NLO
EW corrections are known for the charged-current~\cite{Zykunov:2001mn,
  Dittmaier:2001ay, Baur:2004ig, CarloniCalame:2006zq} as well as the
neutral-current
process~\cite{Baur:1997wa,CarloniCalame:2007cd,Dittmaier:2009cr} and
the predictions have been refined in various ways, e.g.\ upon including
multi-photon radiation~\cite{CarloniCalame:2003ux,Placzek:2003zg,
  CarloniCalame:2006zq,Brensing:2007qm,Dittmaier:2009cr},
photon-induced processes~\cite{DKLH,Arbuzov:2007kp,Brensing:2007qm,
  CarloniCalame:2007cd,Dittmaier:2009cr}, and EW corrections within
the MSSM~\cite{Brensing:2007qm,Dittmaier:2009cr}. Also the interplay
of QCD and EW effects has been investigated~\cite{Cao:2004yy}.

The cross sections for
$\PW/\PZ+1\,\mathrm{jet}$~\cite{Giele:1993dj,Campbell:2002tg,vanderBij:1988ac}
and $\PW/\PZ+2\,\mathrm{jets}$~\cite{Campbell:2002tg} production at NLO
QCD are known for a long time. Recently, NLO QCD results for
$\PW/\PZ+3\, \mathrm{jets}$ and even $\PW+4\, \mathrm{jets}$ production
(in leading-colour approximation) were
presented~\cite{Berger:2010zx}. $\PW/\PZ+1\,\mathrm{jet}$ production has
also been matched with parton showers~\cite{Alioli:2010qp}.  Moreover,
approximate results are available for the NNLO QCD corrections to
$\PZ+\mathrm{jet}/\PZ+2\, \mathrm{jets}$ production for observables with
especially large $K$-factors~\cite{Rubin:2010xp}.  In the EW sector,
higher-order corrections to $\PW+1\,\mathrm{jet}$ production have been
first analyzed in the on-shell
approximation~\cite{Kuhn:2007qc,Kuhn:2007cv,Hollik:2007sq} and later
extended to the full NLO EW corrections for the physical final
state~\cite{Denner:2009gj}, i.e.\ a charged lepton, a neutrino, and a
hard jet.

As far as EW corrections to $\PZ + 1\,\mathrm{jet}$ production are
concerned, only the purely weak one-loop corrections in the SM have been
investigated in the on-shell
approximation~\cite{Kuhn:2004em,Kuhn:2005az}, i.e.\ in this calculation
the \PZ\ boson is treated as a stable external particle and photonic
corrections have been ignored.  For \PZ\ bosons at large transverse
momentum, requiring a large centre-of-mass energy, using on-shell \PZ\
bosons is a good approximation since the EW corrections are dominated by
large universal Sudakov logarithms~\cite{Ciafaloni:1998xg}. In
\citere{Kuhn:2004em} the leading corrections up to the next-to-leading
logarithms at the one- and two-loop level have been calculated.  Later
the full NLO weak corrections have been added~\cite{Kuhn:2005az}.
However, the on-shell calculation is limited to a particular kinematic
regime where neither off-shell effects nor the event definition for the
physical final state play a role.  Finally, photonic corrections have to
be taken into account for a precision at the level of a couple of
percent.

In this work, we present a calculation of the NLO (i.e.\ one-loop) EW
corrections for the physical final state, i.e.\ for the process
$\Pp\Pp/\Pp\bar\Pp \to \Pl^+\Pl^- + \mathrm{jet} +\X$. Following our
earlier work~\cite{Denner:2009gj} on $\PW+1\,\mathrm{jet}$ production,
the \PZ-boson resonance is described in the complex-mass
scheme~\cite{Denner:1999gp,Denner:2005fg}. All off-shell effects due to
the finite width of the \PZ\ boson, the contributions of and the
interference with an intermediate photon, and photonic corrections are
included. Our results have been implemented in a fully flexible Monte
Carlo code which is able to calculate binned distributions for all
physically relevant $\PZ+1\,\mathrm{jet}$ observables.  In real emission
events with photons inside a jet, we distinguish $\PZ+\mathrm{jet}$ and
$\PZ+\mathrm{photon}$ production by a cut on the photon energy fraction
inside the jet employing a measured quark-to-photon fragmentation
function \cite{Buskulic:1995au}.

Our calculation is completely generic and thus not limited to specific
observables or kinematic regimes. The interplay between the
(potentially resonant) \PZ\ boson and the off-shell photon is included
without approximations at NLO and can also be studied for observables
for which the exchanged \PZ\ boson is far off shell and the photon
contribution is potentially sizeable.  However, we rely on the
presence of a hard jet with sizable transverse momentum in the final
state. For final-state jets which become soft or collinear to the beam
pipe, the calculation breaks down.  Nevertheless, the calculation of
the EW corrections for \PZ\ production in association with a hard jet
is also a step towards the mixed NNLO EW and QCD corrections to
inclusive \PZ-boson production. 

We have also recalculated the NLO QCD corrections at
$\mathcal{O}(\alpha^2 \alpha^2_\mathrm{s})$ in a fully flexible way,
supporting a phase-space dependent choice for the factorization and
renormalization scales, as will be discussed in some detail in
\refse{se:numres}.

This paper is organized as follows. In \refse{se:details}, we describe
our calculation in detail and discuss all the theoretical concepts and
tools which have been used. In \refse{se:numres}, we specify the
numerical input as well as the details of our event selection.
Numerical results are given for $\PZ+\mathrm{jet}$ production both at the 
LHC and at the Tevatron. We present inclusive cross sections for specified 
sets of cuts as well as distributions for the relevant observables.
Our conclusions are given in \refse{se:concl}.

\section{Details of the calculation}
\label{se:details} 

\subsection{General setup}
\label{se:setup} 

At hadron colliders, the production of a charged-lepton pair via \PZ-boson 
or photon exchange in association with one hard jet is governed at leading 
order (LO) by quark--antiquark fusion, where the
initial-state quarks radiate a gluon, and the corresponding crossed
channels with a gluon in the initial state. To be specific,
the relevant partonic processes are
\begin{eqnarray}
\label{eq:proc1}
& \Pq_i \, \, \Pqbar_i & \to \PZ/\gamma^* \Pg      \to \Plp \Plm \, \Pg \, ,\\
\label{eq:proc2}
& \Pq_i \, \, \Pg      & \to \PZ/\gamma^* \Pq_i    \to \Plp \Plm \, \Pq_i \, ,\\
\label{eq:proc3}
& \Pqbar_i \, \, \Pg   & \to \PZ/\gamma^* \Pqbar_i \to \Plp \Plm \, \Pqbar_i \, ,
\end{eqnarray}
where $\Pq_i$ denotes any light quark, i.e.\ 
$\Pq_i=\Pu,\Pd,\Pc,\Ps,\Pb$.  The corresponding tree-level Feynman
diagrams for process \refeq{eq:proc1} are shown in
\reffi{fi:born_udWg}. The intermediate \PZ-boson resonance is
described by a complex \PZ-boson mass $ \mu_{\PZ}$ via the replacement
\bfi
\begin{center}
\unitlength=3.bp%

\begin{center}
\begin{small}
\begin{feynartspicture}(80,30)(2,1)

\FADiagram{}
\FAProp(0.,15.)(10.,13.)(0.,){/Straight}{1}
\FALabel(5.30398,15.0399)[b]{$q$}
\FAProp(0.,5.)(10.,5.5)(0.,){/Straight}{-1}
\FALabel(5.0774,4.18193)[t]{$q$}
\FAProp(20.,17.)(15.5,13.5)(0.,){/Straight}{-1}
\FALabel(17.2784,15.9935)[br]{$l$}
\FAProp(20.,10.)(15.5,13.5)(0.,){/Straight}{1}
\FALabel(18.2216,12.4935)[bl]{$l$}
\FAProp(20.,3.)(10.,5.5)(0.,){/Cycles}{0}
\FALabel(15.3759,5.27372)[b]{$\mathrm{g}$}
\FAProp(10.,13.)(10.,5.5)(0.,){/Straight}{1}
\FALabel(8.93,9.25)[r]{$q$}
\FAProp(10.,13.)(15.5,13.5)(0.,){/Sine}{0}
\FALabel(11.8713,16.8146)[t]{$\mathrm{Z}/\gamma$}
\FAVert(10.,13.){0}
\FAVert(10.,5.5){0}
\FAVert(15.5,13.5){0}

\FADiagram{}
\FAProp(0.,15.)(10.,5.5)(0.,){/Straight}{1}
\FALabel(3.19219,13.2012)[bl]{$q$}
\FAProp(0.,5.)(10.,13.)(0.,){/Straight}{-1}
\FALabel(3.17617,6.28478)[tl]{$q$}
\FAProp(20.,17.)(16.,13.5)(0.,){/Straight}{-1}
\FALabel(18.4593,15.9365)[br]{$l$}
\FAProp(20.,10.)(16.,13.5)(0.,){/Straight}{1}
\FALabel(17.4593,11.0635)[tr]{$l$}
\FAProp(20.,3.)(10.,5.5)(0.,){/Cycles}{0}
\FALabel(14.4543,2.54718)[t]{$\mathrm{g}$}
\FAProp(10.,5.5)(10.,13.)(0.,){/Straight}{1}
\FALabel(11.07,9.25)[l]{$q$}
\FAProp(10.,13.)(16.,13.5)(0.,){/Sine}{0}
\FALabel(12.8713,14.3146)[b]{$\mathrm{Z}/\gamma$}
\FAVert(10.,5.5){0}
\FAVert(10.,13.){0}
\FAVert(16.,13.5){0}

\end{feynartspicture}
\end{small}
\end{center}

\vspace*{-1.6em}
\end{center}
\mycaption{\label{fi:born_udWg} Feynman diagrams for the LO process \refeq{eq:proc1}.}
\efi
\begin{equation}
\MZ^2 \to \mu_{\PZ}^2 = \MZ^2 -\ri \MZ \GZ
\end{equation}
in the \PZ~propagator as dictated by the complex-mass scheme (see
below). Hence, all our results correspond to a fixed-width description
of the Breit--Wigner resonance. Moreover, all related quantities,
in particular the weak mixing angle, have to be formulated in terms of
the complex mass parameters. The final-state leptons are 
treated as massless unless their small masses are used to regularize a 
collinear divergence.

The tree-level amplitudes do not depend on the quark generation and only
differ for up- and down-type quarks due to the different quantum numbers
in the $\PZ\Pq\Pqbar$ vertex. Hence, the summation over the quark
flavours is straightforward for each of the three process types shown in
\refeq{eq:proc1}--\refeq{eq:proc3} when folding the squared tree-level
amplitudes with the corresponding PDFs.  The five quark flavours
(including the bottom quark), which appear as external particles, are
treated as massless throughout the calculation, except if small masses
are used to regularize a collinear divergence. At tree level the
bottom-quark-induced processes do not show any peculiarities. Only in
the evaluation of the EW virtual corrections they have to be treated
with special care (see \refse{se:virt}).

In this work, we describe $\PZ+\mathrm{jet}$ production at NLO
accuracy w.r.t.\ EW  contributions, i.e.\ at the order
$\mathcal{O}(\alpha^3 \alpha_{\mathrm{s}})$. Hence, we also include the tree-level 
processes with a photon in the initial state,
\begin{eqnarray}
\label{eq:proc4}
& \Pq_i \, \, \ga      & \to \PZ/\gamma^* \Pq_i    \to \Plp \Plm \, \Pq_i \, ,\\
\label{eq:proc5}
& \Pqbar_i \, \, \ga   & \to \PZ/\gamma^* \Pqbar_i \to \Plp \Plm \, \Pqbar_i \, ,
\end{eqnarray}
which contribute at the order $\mathcal{O}(\alpha^3)$ and may thus lead
to relevant corrections at the expected accuracy level of a few percent.
\bfi
\begin{center}
\unitlength=3.bp%

\begin{center}
\begin{small}
\begin{feynartspicture}(150,30)(4,1)

\FADiagram{}
\FAProp(0.,15.)(5.5,10.)(0.,){/Straight}{1}
\FALabel(2.18736,11.8331)[tr]{$q$}
\FAProp(0.,5.)(5.5,10.)(0.,){/Sine}{0}
\FALabel(3.31264,6.83309)[tl]{$\gamma$}
\FAProp(20.,17.)(15.5,13.5)(0.,){/Straight}{-1}
\FALabel(17.2784,15.9935)[br]{$l$}
\FAProp(20.,10.)(15.5,13.5)(0.,){/Straight}{1}
\FALabel(18.2216,12.4935)[bl]{$l$}
\FAProp(20.,3.)(12.,10.)(0.,){/Straight}{-1}
\FALabel(15.4593,5.81351)[tr]{$q$}
\FAProp(5.5,10.)(12.,10.)(0.,){/Straight}{1}
\FALabel(8.75,8.93)[t]{$q$}
\FAProp(15.5,13.5)(12.,10.)(0.,){/Sine}{0}
\FALabel(13.134,12.366)[br]{$\mathrm{Z}/\gamma$}
\FAVert(5.5,10.){0}
\FAVert(15.5,13.5){0}
\FAVert(12.,10.){0}

\FADiagram{}
\FAProp(0.,15.)(10.,5.5)(0.,){/Straight}{1}
\FALabel(3.19219,13.2012)[bl]{$q$}
\FAProp(0.,5.)(10.,14.5)(0.,){/Sine}{0}
\FALabel(3.10297,6.58835)[tl]{$\gamma$}
\FAProp(20.,17.)(10.,14.5)(0.,){/Straight}{-1}
\FALabel(14.6241,16.7737)[b]{$l$}
\FAProp(20.,10.)(10.,10.)(0.,){/Straight}{1}
\FALabel(15.95,11.07)[b]{$l$}
\FAProp(20.,3.)(10.,5.5)(0.,){/Straight}{-1}
\FALabel(14.6241,3.22628)[t]{$q$}
\FAProp(10.,5.5)(10.,10.)(0.,){/Sine}{0}
\FALabel(11.07,7.75)[l]{$\mathrm{Z}/\gamma$}
\FAProp(10.,14.5)(10.,10.)(0.,){/Straight}{-1}
\FALabel(11.07,12.25)[l]{$l$}
\FAVert(10.,5.5){0}
\FAVert(10.,14.5){0}
\FAVert(10.,10.){0}

\FADiagram{}
\FAProp(0.,15.)(10.,14.5)(0.,){/Straight}{1}
\FALabel(5.0774,15.8181)[b]{$q$}
\FAProp(0.,5.)(10.,5.5)(0.,){/Sine}{0}
\FALabel(5.0774,4.18193)[t]{$\gamma$}
\FAProp(20.,17.)(10.,10.)(0.,){/Straight}{-1}
\FALabel(15.8366,15.2248)[br]{$l$}
\FAProp(20.,10.)(10.,5.5)(0.,){/Straight}{1}
\FALabel(17.7693,10.1016)[br]{$l$}
\FAProp(20.,3.)(10.,14.5)(0.,){/Straight}{-1}
\FALabel(16.7913,4.81596)[tr]{$q$}
\FAProp(10.,14.5)(10.,10.)(0.,){/Sine}{0}
\FALabel(8.93,12.25)[r]{$\mathrm{Z}/\gamma$}
\FAProp(10.,5.5)(10.,10.)(0.,){/Straight}{1}
\FALabel(8.93,7.75)[r]{$l$}
\FAVert(10.,14.5){0}
\FAVert(10.,5.5){0}
\FAVert(10.,10.){0}

\FADiagram{}
\FAProp(0.,15.)(10.,13.)(0.,){/Straight}{1}
\FALabel(5.30398,15.0399)[b]{$q$}
\FAProp(0.,5.)(10.,5.5)(0.,){/Sine}{0}
\FALabel(5.0774,4.18193)[t]{$\gamma$}
\FAProp(20.,17.)(15.5,13.5)(0.,){/Straight}{-1}
\FALabel(17.2784,15.9935)[br]{$l$}
\FAProp(20.,10.)(15.5,13.5)(0.,){/Straight}{1}
\FALabel(18.2216,12.4935)[bl]{$l$}
\FAProp(20.,3.)(10.,5.5)(0.,){/Straight}{-1}
\FALabel(15.3759,5.27372)[b]{$q$}
\FAProp(10.,13.)(10.,5.5)(0.,){/Straight}{1}
\FALabel(8.93,9.25)[r]{$q$}
\FAProp(10.,13.)(15.5,13.5)(0.,){/Sine}{0}
\FALabel(12.8903,12.1864)[t]{$\mathrm{Z}/\gamma$}
\FAVert(10.,13.){0}
\FAVert(10.,5.5){0}
\FAVert(15.5,13.5){0}

\end{feynartspicture}
\end{small}
\end{center}

\vspace*{-1.6em}
\end{center}
\mycaption{\label{fi:born_ugaWd} 
Feynman diagrams for the photon-induced process \refeq{eq:proc4}.}
\efi
The tree-level Feynman diagrams for process \refeq{eq:proc4} are shown
in \reffi{fi:born_ugaWd}. The photon content of the proton has been
quantified in the MRSTQED2004 PDFs~\cite{Martin:2004dh}.  Since the
photon also couples to the charged leptons in the final state, the
amplitude is more involved than its QCD counterpart. As in our earlier
work on $\PW+\mathrm{jet}$ production, in this work, we do not
consider the crossed processes corresponding to $\Pl^+ \Pl^-
+\mathrm{photon}$ production, which would lead to tiny corrections.
The non-trivial definition and separation of the $\Pl^+ \Pl^-
+\mathrm{jet}$ and $\Pl^+ \Pl^- +\mathrm{photon}$ final states when
additional photons are present due to bremsstrahlung are discussed in
detail in \refse{se:real}.

To complete the description of the general setup of our calculation,
we also repeat a few points which do not differ from our earlier
calculation for $\PW+\mathrm{jet}$ production. To define the
electromagnetic coupling constant $\alpha$, we use the $\GF$ scheme,
\ie we derive $\alpha$ from the Fermi constant according to
\beq
 \alpha_{\GF} = \frac{\sqrt{2}\GF\MW^2}{\pi}\left(1-\frac{\MW^2}{\MZ^2}\right).
\eeq
In this scheme, the weak corrections to muon decay $\De r$ are
included in the charge renormalization constant (see \eg
\citere{Dittmaier:2001ay}).  As a consequence, the EW corrections are
independent of logarithms of the light-quark masses. Moreover, this
definition effectively resums the contributions associated with the
running of $\al$ from zero to the weak scale and absorbs some leading
universal corrections $\propto\GF\Mt^2$ from the $\rho$~parameter into
the LO amplitude.

For corrections due to collinear final-state radiation it would be
more appropriate to use $\alpha(0)$ defined in the Thomson limit to
describe the corresponding coupling. On the other hand, using
$\alpha_{\GF}$ everywhere is best suited to describe the large
corrections due to Sudakov logarithms in the high-energy regime. Thus,
the optimal choice cannot be achieved in one particular input scheme,
and necessarily the calculation requires more refinements beyond NLO.
In particular, among other things, higher-order effects from
multi-photon emission should also be included at this level of
precision which is beyond the scope of this work. We find that the
difference of the two schemes in an NLO calculation only amounts to
about 3\% of the EW corrections.

We employ the traditional Feynman-diagrammatic approach to calculate
all relevant amplitudes in the 't Hooft--Feynman gauge. For a
numerical evaluation at the amplitude level we use the
Weyl--van-der-Waerden spinor formalism.  To ensure the correctness of
the presented results we have performed two independent calculations
which are in mutual agreement.

One calculation starts from diagrammatic expressions for the one-loop
corrections generated by {\sc FeynArts} 1.0 \cite{Kublbeck:1990xc}.
The algebraic evaluation of the loop amplitudes is performed with an 
in-house program written in {\sl Mathematica}, and the results are 
automatically transferred to {\sl Fortran}. 
The Born and bremsstrahlung amplitudes are
calculated and optimized by hand and directly included into a {\sl Fortran}
program for numerical evaluation. A specific parametrization of phase
space is used for an adaptive Monte Carlo integration employing the
{\sc Vegas} \cite{Lepage:1977sw} algorithm.

The second calculation is based on {\sc FeynArts} 3.2 \cite{Hahn:2000kx}
and {\sc FormCalc} version 3.1 \cite{Hahn:1998yk}.  The translation of
the amplitudes into the Weyl--van-der-Waerden formalism as presented in
\citere{Dittmaier:1998nn} is performed with the program {\sc
  Pole}~\cite{Accomando:2005ra}.  {\sc Pole} also provides an interface
to the multi-channel phase-space integrator {\sc Lusifer}
\cite{Dittmaier:2002ap} which has been extended to use {\sc Vegas}
in order to optimize each phase-space
mapping. {\sc Madgraph}~\cite{Alwall:2007st} has been very useful for internal
checks of the real-emission amplitudes.

\subsection{Virtual corrections}
\label{se:virt} 

We calculate the virtual one-loop QCD and EW corrections for the
partonic processes \refeq{eq:proc1}--\refeq{eq:proc3}, but do not
include the NLO QCD corrections to the photon-induced processes which
are formally part of the corrections up to $\mathcal{O}(\alpha^3
\alpha_{\mathrm{s}})$.  The cross section of the photon-induced
processes turns out to be numerically small at LO, and the corresponding
corrections in the case of $\PW+\mathrm{jet}$ production turned out to
be completely negligible. Hence, we do not expect phenomenologically
relevant corrections for $\PZ+\mathrm{jet}$ either. We do not include
the (loop-induced) contributions of the partonic process
$\Pg\Pg\to\PZ\,\Pg$. Based on \citere{vanderBij:1988ac} its contribution
at the LHC and the Tevatron can be estimated to be below one percent.

The calculated virtual QCD corrections are straightforward to
implement and consist of up to box (4-point) diagrams only. The NLO EW
corrections are more involved and additionally include pentagon
(5-point) diagrams.  There are $\mathcal{O}(200)$ diagrams per
partonic channel, including 9 pentagons and 32 boxes. The general
structure of the contributions is completely equivalent to our earlier
work on $\PW+\mathrm{jet}$ production. However, $\PZ+\mathrm{jet}$ is
computationally more demanding, not only because there are more
diagrams per partonic channel but also because there are more helicity
combinations contributing to the cross section which are summed at
each phase-space point.  The generic structure of the contributing
diagrams is indicated in \reffi{fi:EW_VF_udWg}, and the pentagon
diagrams are explicitly given in \reffi{fi:EW_pent_udWg}.  The
different channels are related by crossing symmetry.

Concerning the EW corrections, the partonic processes with one or two
(anti-) bottom quarks in the initial state play a special role: There
are box and pentagon diagrams with two $\PW$ bosons and a heavy top
quark in the loop. For all other partonic processes, only massless
quarks propagate in the loops because we neglect CKM mixing, i.e.\ we
set the CKM matrix to unity in our calculation. The mass of the top
quark shifts the total EW correction roughly by a per mille for the
most inclusive cross section discussed in \refse{se:numres}.

\bfi
\begin{center}
\unitlength=2.5bp%

\begin{feynartspicture}(160,40)(4,1)
\FALabel(-5,20)[l]{Self-energy insertions:}
\FADiagram{}
\FAProp(0.,15.)(4.,10.)(0.,){/Straight}{1}
\FALabel(1.26965,12.0117)[tr]{$q$}
\FAProp(0.,5.)(4.,10.)(0.,){/Cycles}{0}
\FALabel(2.73035,7.01172)[tl]{$\mathrm{g}$}
\FAProp(20.,17.)(15.5,13.5)(0.,){/Straight}{-1}
\FALabel(17.2784,15.9935)[br]{$l$}
\FAProp(20.,10.)(15.5,13.5)(0.,){/Straight}{1}
\FALabel(18.2216,12.4935)[bl]{$l$}
\FAProp(20.,3.)(12.,10.)(0.,){/Straight}{-1}
\FALabel(16.5407,7.18649)[bl]{$\text{q}$}
\FAProp(8.,10.)(4.,10.)(0.,){/Straight}{-1}
\FALabel(6.,11.07)[b]{$q$}
\FAProp(8.,10.)(12.,10.)(0.,){/Straight}{1}
\FALabel(10.,8.93)[t]{$q$}
\FAProp(15.5,13.5)(12.,10.)(0.,){/Sine}{0}
\FALabel(13.134,12.366)[br]{$\mathrm{Z/\gamma}$}
\FAVert(4.,10.){0}
\FAVert(15.5,13.5){0}
\FAVert(12.,10.){0}
\FAVert(8.,10.){-1}

\FADiagram{}
\FAProp(0.,15.)(3.5,10.)(0.,){/Straight}{1}
\FALabel(0.960191,12.0911)[tr]{$q$}
\FAProp(0.,5.)(3.5,10.)(0.,){/Cycles}{0}
\FALabel(2.53981,7.09113)[tl]{$\mathrm{g}$}
\FAProp(20.,17.)(15.5,13.5)(0.,){/Straight}{-1}
\FALabel(17.2784,15.9935)[br]{$l$}
\FAProp(20.,10.)(15.5,13.5)(0.,){/Straight}{1}
\FALabel(18.2216,12.4935)[bl]{$l$}
\FAProp(20.,3.)(9.,10.)(0.,){/Straight}{-1}
\FALabel(14.1478,5.67232)[tr]{$\text{q}$}
\FAProp(12.25,11.75)(15.5,13.5)(0.,){/Sine}{0}
\FALabel(14.1299,11.7403)[tl]{$\mathrm{Z/\gamma}$}
\FAProp(12.25,11.75)(9.,10.)(0.,){/Sine}{0}
\FALabel(10.3701,11.7597)[br]{$\mathrm{Z/\gamma}$}
\FAProp(3.5,10.)(9.,10.)(0.,){/Straight}{1}
\FALabel(6.25,8.93)[t]{$q$}
\FAVert(3.5,10.){0}
\FAVert(15.5,13.5){0}
\FAVert(9.,10.){0}
\FAVert(12.25,11.75){-1}

\FADiagram{}
\FAProp(0.,15.)(10.,13.5)(0.,){/Straight}{1}
\FALabel(5.22993,15.3029)[b]{$q$}
\FAProp(0.,5.)(10.,4.5)(0.,){/Cycles}{0}
\FALabel(4.9226,3.68193)[t]{$\mathrm{g}$}
\FAProp(20.,17.)(16.,13.5)(0.,){/Straight}{-1}
\FALabel(17.4593,15.9365)[br]{$l$}
\FAProp(20.,10.)(16.,13.5)(0.,){/Straight}{1}
\FALabel(18.5407,12.4365)[bl]{$l$}
\FAProp(20.,3.)(10.,4.5)(0.,){/Straight}{-1}
\FALabel(15.2299,4.80285)[b]{$\text{q}$}
\FAProp(10.,9.)(10.,13.5)(0.,){/Straight}{-1}
\FALabel(8.93,11.25)[r]{$\text{q}$}
\FAProp(10.,9.)(10.,4.5)(0.,){/Straight}{1}
\FALabel(8.93,6.75)[r]{$\text{q}$}
\FAProp(10.,13.5)(16.,13.5)(0.,){/Sine}{0}
\FALabel(13.,14.57)[b]{$\mathrm{Z/\gamma}$}
\FAVert(10.,13.5){0}
\FAVert(10.,4.5){0}
\FAVert(16.,13.5){0}
\FAVert(10.,9.){-1}

\FADiagram{}
\FAProp(0.,15.)(9.,13.5)(0.,){/Straight}{1}
\FALabel(4.75482,15.2989)[b]{$q$}
\FAProp(0.,5.)(9.,5.5)(0.,){/Cycles}{0}
\FALabel(4.58598,4.18239)[t]{$\mathrm{g}$}
\FAProp(20.,17.)(16.,13.5)(0.,){/Straight}{-1}
\FALabel(17.4593,15.9365)[br]{$l$}
\FAProp(20.,10.)(16.,13.5)(0.,){/Straight}{1}
\FALabel(18.5407,12.4365)[bl]{$l$}
\FAProp(20.,3.)(9.,5.5)(0.,){/Straight}{-1}
\FALabel(14.8435,5.28146)[b]{$\text{q}$}
\FAProp(12.5,13.5)(9.,13.5)(0.,){/Sine}{0}
\FALabel(10.75,15.57)[b]{$\mathrm{Z/\gamma}$}
\FAProp(12.5,13.5)(16.,13.5)(0.,){/Sine}{0}
\FALabel(14.25,11.43)[t]{$\mathrm{Z/\gamma}$}
\FAProp(9.,13.5)(9.,5.5)(0.,){/Straight}{1}
\FALabel(7.93,9.5)[r]{$\text{q}$}
\FAVert(9.,13.5){0}
\FAVert(9.,5.5){0}
\FAVert(16.,13.5){0}
\FAVert(12.5,13.5){-1}
\end{feynartspicture}

\begin{feynartspicture}(120,80)(3,2)
\FALabel(-16,42.5)[l]{Triangle insertions:}
\FADiagram{}
\FAProp(0.,15.)(5.5,10.)(0.,){/Straight}{1}
\FALabel(2.18736,11.8331)[tr]{$q$}
\FAProp(0.,5.)(5.5,10.)(0.,){/Cycles}{0}
\FALabel(3.31264,6.83309)[tl]{$\mathrm{g}$}
\FAProp(20.,17.)(15.5,13.5)(0.,){/Straight}{-1}
\FALabel(17.2784,15.9935)[br]{$l$}
\FAProp(20.,10.)(15.5,13.5)(0.,){/Straight}{1}
\FALabel(18.2216,12.4935)[bl]{$l$}
\FAProp(20.,3.)(12.,10.)(0.,){/Straight}{-1}
\FALabel(15.4593,5.81351)[tr]{$\text{q}$}
\FAProp(5.5,10.)(12.,10.)(0.,){/Straight}{1}
\FALabel(8.75,8.93)[t]{$q$}
\FAProp(15.5,13.5)(12.,10.)(0.,){/Sine}{0}
\FALabel(13.134,12.366)[br]{$\mathrm{Z/\gamma}$}
\FAVert(15.5,13.5){0}
\FAVert(12.,10.){0}
\FAVert(5.5,10.){-1}

\FADiagram{}
\FAProp(0.,15.)(5.5,10.)(0.,){/Straight}{1}
\FALabel(2.18736,11.8331)[tr]{$q$}
\FAProp(0.,5.)(5.5,10.)(0.,){/Cycles}{0}
\FALabel(3.31264,6.83309)[tl]{$\mathrm{g}$}
\FAProp(20.,17.)(15.5,13.5)(0.,){/Straight}{-1}
\FALabel(17.2784,15.9935)[br]{$l$}
\FAProp(20.,10.)(15.5,13.5)(0.,){/Straight}{1}
\FALabel(18.2216,12.4935)[bl]{$l$}
\FAProp(20.,3.)(12.,10.)(0.,){/Straight}{-1}
\FALabel(15.4593,5.81351)[tr]{$\text{q}$}
\FAProp(5.5,10.)(12.,10.)(0.,){/Straight}{1}
\FALabel(8.75,8.93)[t]{$q$}
\FAProp(15.5,13.5)(12.,10.)(0.,){/Sine}{0}
\FALabel(13.134,12.366)[br]{$\mathrm{Z/\gamma}$}
\FAVert(5.5,10.){0}
\FAVert(15.5,13.5){0}
\FAVert(12.,10.){-1}

\FADiagram{}
\FAProp(0.,15.)(5.5,10.)(0.,){/Straight}{1}
\FALabel(2.18736,11.8331)[tr]{$q$}
\FAProp(0.,5.)(5.5,10.)(0.,){/Cycles}{0}
\FALabel(3.31264,6.83309)[tl]{$\mathrm{g}$}
\FAProp(20.,17.)(15.5,13.5)(0.,){/Straight}{-1}
\FALabel(17.2784,15.9935)[br]{$l$}
\FAProp(20.,10.)(15.5,13.5)(0.,){/Straight}{1}
\FALabel(18.2216,12.4935)[bl]{$l$}
\FAProp(20.,3.)(12.,10.)(0.,){/Straight}{-1}
\FALabel(15.4593,5.81351)[tr]{$\text{q}$}
\FAProp(5.5,10.)(12.,10.)(0.,){/Straight}{1}
\FALabel(8.75,8.93)[t]{$q$}
\FAProp(15.5,13.5)(12.,10.)(0.,){/Sine}{0}
\FALabel(13.134,12.366)[br]{$\mathrm{Z/\gamma}$}
\FAVert(5.5,10.){0}
\FAVert(12.,10.){0}
\FAVert(15.5,13.5){-1}


\FADiagram{}
\FAProp(0.,15.)(10.,13.)(0.,){/Straight}{1}
\FALabel(5.30398,15.0399)[b]{$q$}
\FAProp(0.,5.)(10.,5.5)(0.,){/Cycles}{0}
\FALabel(5.0774,4.18193)[t]{$\mathrm{g}$}
\FAProp(20.,17.)(15.5,13.5)(0.,){/Straight}{-1}
\FALabel(17.2784,15.9935)[br]{$l$}
\FAProp(20.,10.)(15.5,13.5)(0.,){/Straight}{1}
\FALabel(18.2216,12.4935)[bl]{$l$}
\FAProp(20.,3.)(10.,5.5)(0.,){/Straight}{-1}
\FALabel(15.3759,5.27372)[b]{$\text{q}$}
\FAProp(10.,13.)(10.,5.5)(0.,){/Straight}{1}
\FALabel(8.93,9.25)[r]{$\text{q}$}
\FAProp(10.,13.)(15.5,13.5)(0.,){/Sine}{0}
\FALabel(12.8903,12.1864)[t]{$\mathrm{Z/\gamma}$}
\FAVert(10.,13.){0}
\FAVert(15.5,13.5){0}
\FAVert(10.,5.5){-1}



\FADiagram{}
\FAProp(0.,15.)(10.,13.)(0.,){/Straight}{1}
\FALabel(5.30398,15.0399)[b]{$q$}
\FAProp(0.,5.)(10.,5.5)(0.,){/Cycles}{0}
\FALabel(5.0774,4.18193)[t]{$\mathrm{g}$}
\FAProp(20.,17.)(15.5,13.5)(0.,){/Straight}{-1}
\FALabel(17.2784,15.9935)[br]{$l$}
\FAProp(20.,10.)(15.5,13.5)(0.,){/Straight}{1}
\FALabel(18.2216,12.4935)[bl]{$l$}
\FAProp(20.,3.)(10.,5.5)(0.,){/Straight}{-1}
\FALabel(15.3759,5.27372)[b]{$\text{q}$}
\FAProp(10.,13.)(10.,5.5)(0.,){/Straight}{1}
\FALabel(8.93,9.25)[r]{$\text{q}$}
\FAProp(10.,13.)(15.5,13.5)(0.,){/Sine}{0}
\FALabel(12.8903,12.1864)[t]{$\mathrm{Z/\gamma}$}
\FAVert(10.,5.5){0}
\FAVert(15.5,13.5){0}
\FAVert(10.,13.){-1}

\FADiagram{}
\FAProp(0.,15.)(10.,13.)(0.,){/Straight}{1}
\FALabel(5.30398,15.0399)[b]{$q$}
\FAProp(0.,5.)(10.,5.5)(0.,){/Cycles}{0}
\FALabel(5.0774,4.18193)[t]{$\mathrm{g}$}
\FAProp(20.,17.)(15.5,13.5)(0.,){/Straight}{-1}
\FALabel(17.2784,15.9935)[br]{$l$}
\FAProp(20.,10.)(15.5,13.5)(0.,){/Straight}{1}
\FALabel(18.2216,12.4935)[bl]{$l$}
\FAProp(20.,3.)(10.,5.5)(0.,){/Straight}{-1}
\FALabel(15.3759,5.27372)[b]{$\text{q}$}
\FAProp(10.,13.)(10.,5.5)(0.,){/Straight}{1}
\FALabel(8.93,9.25)[r]{$\text{q}$}
\FAProp(10.,13.)(15.5,13.5)(0.,){/Sine}{0}
\FALabel(12.8903,12.1864)[t]{$\mathrm{Z/\gamma}$}
\FAVert(10.,13.){0}
\FAVert(10.,5.5){0}
\FAVert(15.5,13.5){-1}

\end{feynartspicture}

\begin{feynartspicture}(160,40)(4,1)
\FALabel(-5,22)[l]{Box and pentagon insertions:}
\FADiagram{}
\FAProp(0.,15.)(5.5,10.)(0.,){/Straight}{1}
\FALabel(2.18736,11.8331)[tr]{$q$}
\FAProp(0.,5.)(5.5,10.)(0.,){/Cycles}{0}
\FALabel(3.31264,6.83309)[tl]{$\mathrm{g}$}
\FAProp(20.,17.)(12,10)(0.,){/Straight}{-1}
\FALabel(17.2784,15.9935)[br]{$l$}
\FAProp(20.,10.)(12,10)(0.,){/Straight}{1}
\FALabel(18.2216,10.4935)[bl]{$l$}
\FAProp(20.,3.)(12.,10.)(0.,){/Straight}{-1}
\FALabel(15.4593,5.81351)[tr]{$\text{q}$}
\FAProp(5.5,10.)(12.,10.)(0.,){/Straight}{1}
\FALabel(8.75,8.93)[t]{$q$}
\FAVert(5.5,10.){0}
\FAVert(12.,10.){-1}

\FADiagram{}
\FAProp(0.,15.)(10.,13.)(0.,){/Straight}{1}
\FALabel(5.30398,15.0399)[b]{$q$}
\FAProp(0.,5.)(10.,5.5)(0.,){/Cycles}{0}
\FALabel(5.0774,4.18193)[t]{$\mathrm{g}$}
\FAProp(20.,17.)(10.,13.)(0.,){/Straight}{-1}
\FALabel(17.2784,15.9935)[br]{$l$}
\FAProp(20.,10.)(10,13.)(0.,){/Straight}{1}
\FALabel(17.2216,11.4935)[bl]{$l$}
\FAProp(20.,3.)(10.,5.5)(0.,){/Straight}{-1}
\FALabel(15.3759,5.27372)[b]{$\text{q}$}
\FAProp(10.,13.)(10.,5.5)(0.,){/Straight}{1}
\FALabel(8.93,9.25)[r]{$\text{q}$}
\FAVert(10.,13.){0}
\FAVert(10.,5.5){0}
\FAVert(10.,13.){-1}

\FADiagram{}
\FAProp(0.,15.)(10,10.)(0.,){/Straight}{1}
\FALabel(4.18736,11.8331)[tr]{$q$}
\FAProp(0.,5.)(10,10.)(0.,){/Cycles}{0}
\FALabel(3.31264,5.83309)[tl]{$\mathrm{g}$}
\FAProp(20.,17.)(15.5,13.5)(0.,){/Straight}{-1}
\FALabel(17.2784,15.9935)[br]{$l$}
\FAProp(20.,10.)(15.5,13.5)(0.,){/Straight}{1}
\FALabel(18.2216,12.4935)[bl]{$l$}
\FAProp(20.,3.)(10,10.)(0.,){/Straight}{-1}
\FALabel(13,5)[tl]{$\text{q}$}
\FAProp(15.5,13.5)(10,10.)(0.,){/Sine}{0}
\FALabel(13.134,12.366)[br]{$\mathrm{Z/\gamma}$}
\FAVert(15.5,13.5){0}
\FAVert(10,10.){-1}

\FADiagram{}
\FAProp(0.,15.)(10,10.)(0.,){/Straight}{1}
\FALabel(4.18736,11.8331)[tr]{$q$}
\FAProp(0.,5.)(10,10.)(0.,){/Cycles}{0}
\FALabel(3.31264,5.83309)[tl]{$\mathrm{g}$}
\FAProp(20,3)(10,10)(0,){/Straight}{-1}
\FALabel(13,5)[tl]{$\text{q}$}
\FAProp(20.,17.)(10,10)(0.,){/Straight}{-1}
\FALabel(17.4593,15.9365)[br]{$l$}
\FAProp(20.,10.)(10,10)(0.,){/Straight}{1}
\FALabel(17.5407,10.4365)[bl]{$l$}
\FAVert(10.,10.){-1}

\end{feynartspicture}

%


\vspace*{-2em}
\end{center}
\mycaption{\label{fi:EW_VF_udWg} Contributions of different
  one-particle irreducible vertex functions (indicated as blobs)
  to the LO process
  \refeq{eq:proc2}; there are contributions from self-energies,
  triangles, boxes, and pentagon graphs.}
\efi

\bfi
\begin{center}
\unitlength=3.bp%
\begin{small}
\begin{feynartspicture}(120,40)(3,1)

\FADiagram{}
\FAProp(0.,15.)(5.,13.)(0.,){/Straight}{1}
\FALabel(3.07566,14.9591)[b]{$q$}
\FAProp(0.,5.)(5.,7.)(0.,){/Cycles}{0}
\FALabel(3.07566,5.04086)[t]{$\mathrm{g}$}
\FAProp(20.,17.)(10.5,14.5)(0.,){/Straight}{-1}
\FALabel(14.8555,16.769)[b]{$l$}
\FAProp(20.,10.)(14.,10.)(0.,){/Straight}{1}
\FALabel(17.,11.07)[b]{$l$}
\FAProp(20.,3.)(10.5,5.5)(0.,){/Straight}{-1}
\FALabel(14.8555,3.23103)[t]{$q$}
\FAProp(5.,13.)(5.,7.)(0.,){/Straight}{1}
\FALabel(3.93,10.)[r]{$q$}
\FAProp(5.,13.)(10.5,14.5)(0.,){/Sine}{0}
\FALabel(7.34217,14.7654)[b]{$\mathrm{Z}/\gamma$}
\FAProp(5.,7.)(10.5,5.5)(0.,){/Straight}{1}
\FALabel(7.34217,5.23462)[t]{$q$}
\FAProp(10.5,14.5)(14.,10.)(0.,){/Straight}{-1}
\FALabel(12.9935,12.7216)[bl]{$l$}
\FAProp(14.,10.)(10.5,5.5)(0.,){/Sine}{0}
\FALabel(12.9935,7.27839)[tl]{$\mathrm{Z}/\gamma$}
\FAVert(5.,13.){0}
\FAVert(5.,7.){0}
\FAVert(10.5,14.5){0}
\FAVert(14.,10.){0}
\FAVert(10.5,5.5){0}

\FADiagram{}
\FAProp(0.,15.)(5.,13.)(0.,){/Straight}{1}
\FALabel(3.07566,14.9591)[b]{$q$}
\FAProp(0.,5.)(5.,7.)(0.,){/Cycles}{0}
\FALabel(3.07566,5.04086)[t]{$\mathrm{g}$}
\FAProp(20.,17.)(14.,10.)(0.,){/Straight}{-1}
\FALabel(18.2902,16.1827)[br]{$l$}
\FAProp(20.,10.)(10.5,14.5)(0.,){/Straight}{1}
\FALabel(18.373,9.75407)[tr]{$l$}
\FAProp(20.,3.)(10.5,5.5)(0.,){/Straight}{-1}
\FALabel(14.8555,3.23103)[t]{$q$}
\FAProp(5.,13.)(5.,7.)(0.,){/Straight}{1}
\FALabel(3.93,10.)[r]{$q$}
\FAProp(5.,13.)(10.5,14.5)(0.,){/Sine}{0}
\FALabel(7.34217,14.7654)[b]{$\mathrm{Z}/\gamma$}
\FAProp(5.,7.)(10.5,5.5)(0.,){/Straight}{1}
\FALabel(7.34217,5.23462)[t]{$q$}
\FAProp(14.,10.)(10.5,14.5)(0.,){/Straight}{-1}
\FALabel(11.5065,11.7784)[tr]{$l$}
\FAProp(14.,10.)(10.5,5.5)(0.,){/Sine}{0}
\FALabel(12.9935,7.27839)[tl]{$\mathrm{Z}/\gamma$}
\FAVert(5.,13.){0}
\FAVert(5.,7.){0}
\FAVert(14.,10.){0}
\FAVert(10.5,14.5){0}
\FAVert(10.5,5.5){0}

\FADiagram{}
\FAProp(0.,15.)(5.,13.)(0.,){/Straight}{1}
\FALabel(3.07566,14.9591)[b]{$q$}
\FAProp(0.,5.)(5.,7.)(0.,){/Cycles}{0}
\FALabel(3.07566,5.04086)[t]{$\mathrm{g}$}
\FAProp(20.,17.)(14.,10.)(0.,){/Straight}{-1}
\FALabel(18.2902,16.1827)[br]{$l$}
\FAProp(20.,10.)(10.5,14.5)(0.,){/Straight}{1}
\FALabel(18.373,9.75407)[tr]{$l$}
\FAProp(20.,3.)(10.5,5.5)(0.,){/Straight}{-1}
\FALabel(14.8555,3.23103)[t]{$q$}
\FAProp(5.,13.)(5.,7.)(0.,){/Straight}{1}
\FALabel(3.93,10.)[r]{$q'$}
\FAProp(5.,13.)(10.5,14.5)(0.,){/Sine}{-1}
\FALabel(7.34217,14.7654)[b]{$\mathrm{W}$}
\FAProp(5.,7.)(10.5,5.5)(0.,){/Straight}{1}
\FALabel(7.34217,5.23462)[t]{$q'$}
\FAProp(14.,10.)(10.5,14.5)(0.,){/Straight}{-1}
\FALabel(11.5065,11.7784)[tr]{$\nu_l$}
\FAProp(14.,10.)(10.5,5.5)(0.,){/Sine}{-1}
\FALabel(12.9935,7.27839)[tl]{$\mathrm{W}$}
\FAVert(5.,13.){0}
\FAVert(5.,7.){0}
\FAVert(14.,10.){0}
\FAVert(10.5,14.5){0}
\FAVert(10.5,5.5){0}
\end{feynartspicture}
\end{small}

\vspace*{-2em}
\end{center}
\mycaption{\label{fi:EW_pent_udWg}
Virtual pentagon contributions to the process \refeq{eq:proc2}. Note that for
external bottom quarks the exchange of two \PW\ bosons leads to diagrams with 
massive top-quark lines ($q'=\mathrm{top}$) in the loop.}
\efi

The potentially resonant \PZ\ bosons require a proper inclusion of the
finite gauge-boson width in the propagators. We use the complex-mass
scheme, which was introduced in \citere{Denner:1999gp} for LO
calculations and generalized to the one-loop level in
\citere{Denner:2005fg}. In this approach the W- and Z-boson masses are
consistently treated as complex quantities, defined as the locations
of the propagator poles in the complex plane.  This leads to complex
couplings and, in particular, a complex weak mixing angle.  The scheme
fully respects all relations that follow from gauge invariance. A
brief description of the complex-mass scheme can also be found in
\citere{Denner:2006ic}.

The amplitudes can be expressed in terms of ``standard matrix
elements'' \cite{Denner:1991kt}, which comprise all
polarization-dependent quantities (spinor chains, polarization
vectors), and invariant coefficients, which contain the tensor
integrals.  The tensor integrals are recursively reduced to master
integrals at the numerical level.  The standard scalar integrals are
evaluated using two in-house {\sl Fortran} libraries which are based
on the results of \citere{Denner:2010tr} for 4-point functions with
complex internal masses and on the methods and results of
\citere{'tHooft:1979xw}.  Results for different regularization schemes
are translated into each other with the method of
\citere{Dittmaier:2003bc}.  Tensor and scalar 5-point functions are
directly expressed in terms of 4-point integrals
\cite{Melrose:1965kb,Denner:2002ii,Denner:2005nn}, while tensor
4-point and 3-point integrals are recursively reduced to scalar
integrals with the Passarino--Veltman algorithm
\cite{Passarino:1979jh}. Although we already find sufficient numerical
stability with this procedure, we additionally apply the dedicated
expansion methods of \citere{Denner:2005nn} in exceptional phase-space
regions where small Gram determinants appear.

UV divergences are regularized dimensionally.  For the infrared (IR),
i.e.\ soft or collinear, divergences we either use pure dimensional
regularization with massless gluons, photons, and fermions (except for
the top quark), or alternatively pure mass regularization with
infinitesimal photon, gluon, and small fermion masses, which are only
kept in the mass-singular logarithms.  When using dimensional
regularization, the rational terms of IR origin are treated as
described in Appendix~A of \citere{Bredenstein:2008zb}; the ones of UV
origin are always automatically included in the amplitude and tensor
integral reduction.

We use an on-shell renormalization prescription for the EW part of the
SM as detailed in \citere{Denner:2005fg} for the complex-mass scheme.
Employing the \GF\ scheme for the definition of the fine-structure
constant $\alpha$, we include $\De r$ in the charge renormalization
constant as mentioned above. The strong coupling constant is
renormalized in the \MSbar\ scheme with five active flavours. Hence,
bottom quarks are included everywhere in the calculation as a massless
quark flavour.

\subsection{Real corrections}
\label{se:real} 

The evaluation of the real corrections has to be done with 
particular care, both for theoretical
consistency as well as to match the experimental observables as closely as 
possible.
Let us first focus on the EW real corrections to the partonic processes
\refeq{eq:proc1}--\refeq{eq:proc3}. The emission of an additional photon leads to
the processes
\begin{eqnarray}
\label{eq:bremsproc1}
& \Pq_i \, \, \Pqbar_i & \to \Plp \Plm \, \Pg \, \gamma \, ,\\
\label{eq:bremsproc2}
& \Pq_i \, \, \Pg      & \to \Plp \Plm \, \Pq_i \, \gamma \, ,\\
\label{eq:bremsproc3}
& \Pqbar_i \, \, \Pg   & \to \Plp \Plm \, \Pqbar_i \, \gamma \, .
\end{eqnarray}
The Feynman diagrams contributing to the process \refeq{eq:bremsproc2}
are shown in \reffi{fi:EW_real_udWg}.  Due to the emission of soft
photons the real corrections include soft singularities which are
cancelled by corresponding contributions in the virtual corrections
independently of the details of the event selection or recombination
procedure. If the photon and the charged leptons/quarks are recombined
into a pseudo-particle (mimicking the start of hadronic or
electromagnetic showers) to form IR-safe observables, all the
remaining singularities arising from collinear photon emission in the
final state also cancel against the corresponding singularities in the
virtual corrections. This requires that all the selection cuts for a
given observable are blind to the distribution of momenta in collinear
lepton--photon configurations. The left-over collinear singularities
due to collinear photon emission off the initial-state quarks are
absorbed by a redefinition of the PDFs. Technically, we use the dipole
subtraction formalism as specified for photon emission in
\citeres{Dittmaier:1999mb,Dittmaier:2008md} to isolate all the
divergences and ensure the numerical cancellation.

Note that only the MRSTQED2004 PDFs properly account for all QED
effects at NLO. However, this PDF set is outdated by now and does not
include many modern PDF developments.  Hence, we only employ the
MRSTQED2004 set to estimate the photon content of the proton and use
more modern sets for all the partonic channels without photons in the
initial state. In a strict sense, with the choice of a modern PDF set
the calculation is not fully consistent (like a NLO QCD calculation
employing a LO PDF set). However, the numerical effect from a proper
inclusion of the NLO EW corrections in the PDF determination is
expected to be small, since this was also the case for the MRSTQED2004
fit when compared to the corresponding fit neglecting QED effects.
The MRSTQED2004 PDF was defined in the DIS scheme so we stick to this
scheme in our calculation.

\bfi
\begin{center}
\unitlength=2.2bp%
\begin{small}
\begin{feynartspicture}(200,80)(5,2)

\FADiagram{}
\FAProp(0.,15.)(10.,5.5)(0.,){/Straight}{1}
\FALabel(3.19219,13.2012)[bl]{$q$}
\FAProp(0.,5.)(10.,14.5)(0.,){/Cycles}{0}
\FALabel(3.19219,6.79875)[tl]{$\mathrm{g}$}
\FAProp(20.,18.)(17.,15.)(0.,){/Straight}{-1}
\FALabel(18.884,18.116)[br]{$l$}
\FAProp(20.,12.)(17.,15.)(0.,){/Straight}{1}
\FALabel(19.116,14.116)[bl]{$l$}
\FAProp(20.,8.)(13.5,14.)(0.,){/Straight}{-1}
\FALabel(16.1787,10.3411)[tr]{$q$}
\FAProp(20.,2.)(10.,5.5)(0.,){/Sine}{0}
\FALabel(14.488,2.76702)[t]{$\gamma$}
\FAProp(10.,5.5)(10.,14.5)(0.,){/Straight}{1}
\FALabel(11.27,8.5)[l]{$q$}
\FAProp(10.,14.5)(13.5,14.)(0.,){/Straight}{1}
\FALabel(11.5692,15.3044)[b]{$q$}
\FAProp(17.,15.)(13.5,14.)(0.,){/Sine}{0}
\FALabel(14.8242,16.0104)[b]{$\mathrm{Z/\gamma}$}
\FAVert(10.,5.5){0}
\FAVert(10.,14.5){0}
\FAVert(17.,15.){0}
\FAVert(13.5,14.){0}
\FADiagram{}
\FAProp(0.,15.)(4.,10.)(0.,){/Straight}{1}
\FALabel(2.73035,12.9883)[bl]{$q$}
\FAProp(0.,5.)(4.,10.)(0.,){/Cycles}{0}
\FALabel(0.723045,8.42556)[br]{$\mathrm{g}$}
\FAProp(20.,18.)(15.5,15.)(0.,){/Straight}{-1}
\FALabel(17.3702,17.3097)[br]{$l$}
\FAProp(20.,12.)(15.5,15.)(0.,){/Straight}{1}
\FALabel(17.3702,12.6903)[tr]{$l$}
\FAProp(20.,8.)(11.5,12.5)(0.,){/Straight}{-1}
\FALabel(15.5048,9.36013)[tr]{$q$}
\FAProp(20.,2.)(8.,9.)(0.,){/Sine}{0}
\FALabel(13.699,4.64114)[tr]{$\gamma$}
\FAProp(4.,10.)(8.,9.)(0.,){/Straight}{1}
\FALabel(5.62407,8.47628)[t]{$q$}
\FAProp(15.5,15.)(11.5,12.5)(0.,){/Sine}{0}
\FALabel(13.1585,14.5844)[br]{$\mathrm{Z/\gamma}$}
\FAProp(11.5,12.5)(8.,9.)(0.,){/Straight}{-1}
\FALabel(9.13398,11.366)[br]{$q$}
\FAVert(4.,10.){0}
\FAVert(15.5,15.){0}
\FAVert(11.5,12.5){0}
\FAVert(8.,9.){0}

\FADiagram{}
\FAProp(0.,15.)(4.,10.)(0.,){/Straight}{1}
\FALabel(1.26965,12.0117)[tr]{$q$}
\FAProp(0.,5.)(4.,10.)(0.,){/Cycles}{0}
\FALabel(2.73035,7.01172)[tl]{$\mathrm{g}$}
\FAProp(20.,18.)(14.,15.)(0.,){/Straight}{-1}
\FALabel(16.7868,17.4064)[br]{$l$}
\FAProp(20.,12.)(14.,15.)(0.,){/Straight}{1}
\FALabel(16.7868,12.5936)[tr]{$l$}
\FAProp(20.,8.)(14.,5.)(0.,){/Straight}{-1}
\FALabel(16.7868,7.40636)[br]{$q$}
\FAProp(20.,2.)(14.,5.)(0.,){/Sine}{0} 
\FALabel(16.7868,2.59364)[tr]{$\gamma$}
\FAProp(4.,10.)(10.,10.)(0.,){/Straight}{1}
\FALabel(7.,11.07)[b]{$q$}
\FAProp(14.,15.)(10.,10.)(0.,){/Sine}{0}
\FALabel(11.2697,12.9883)[br]{$\mathrm{Z/\gamma}$}
\FAProp(14.,5.)(10.,10.)(0.,){/Straight}{-1}
\FALabel(11.2697,7.01172)[tr]{$q$}
\FAVert(4.,10.){0}
\FAVert(14.,15.){0}
\FAVert(14.,5.){0}
\FAVert(10.,10.){0}

\FADiagram{}
\FAProp(0.,15.)(4.,10.)(0.,){/Straight}{1}
\FALabel(2.73035,12.9883)[bl]{$q$}
\FAProp(0.,5.)(4.,10.)(0.,){/Cycles}{0}
\FALabel(0.723045,8.42556)[br]{$\mathrm{g}$}
\FAProp(20.,18.)(17.,9.5)(0.,){/Straight}{-1}
\FALabel(17.7443,15.5832)[r]{$l$}
\FAProp(20.,12.)(12.,12.5)(0.,){/Straight}{1}
\FALabel(14.0156,13.2695)[b]{$l$}
\FAProp(20.,8.)(9.5,6.)(0.,){/Straight}{-1}
\FALabel(13.1468,5.43442)[t]{$q$}
\FAProp(20.,2.)(17.,9.5)(0.,){/Sine}{0}
\FALabel(17.7479,4.40715)[r]{$\gamma$}
\FAProp(4.,10.)(9.5,6.)(0.,){/Straight}{1}
\FALabel(6.31833,7.22646)[tr]{$q$}
\FAProp(17.,9.5)(12.,12.5)(0.,){/Straight}{-1}
\FALabel(14.1825,10.1509)[tr]{$l$}
\FAProp(12.,12.5)(9.5,6.)(0.,){/Sine}{0}
\FALabel(9.78331,9.80642)[r]{$\mathrm{Z/\gamma}$}
\FAVert(4.,10.){0}
\FAVert(17.,9.5){0}
\FAVert(12.,12.5){0}
\FAVert(9.5,6.){0}

\FADiagram{}
\FAProp(0.,15.)(3.5,10.)(0.,){/Straight}{1}
\FALabel(0.960191,12.0911)[tr]{$q$}
\FAProp(0.,5.)(3.5,10.)(0.,){/Cycles}{0}
\FALabel(2.53981,7.09113)[tl]{$\mathrm{g}$}
\FAProp(20.,18.)(11.,7.5)(0.,){/Straight}{-1}
\FALabel(16.7902,15.6827)[br]{$l$}
\FAProp(20.,12.)(16.,6.)(0.,){/Straight}{1}
\FALabel(18.9649,11.9213)[br]{$l$}
\FAProp(20.,8.)(8.,11.)(0.,){/Straight}{-1}
\FALabel(10.7853,11.7292)[b]{$q$}
\FAProp(20.,2.)(16.,6.)(0.,){/Sine}{0}
\FALabel(17.384,3.38398)[tr]{$\gamma$}
\FAProp(3.5,10.)(8.,11.)(0.,){/Straight}{1}
\FALabel(5.41376,11.5331)[b]{$q$}
\FAProp(11.,7.5)(16.,6.)(0.,){/Straight}{-1}
\FALabel(13.0546,5.74537)[t]{$l$}
\FAProp(11.,7.5)(8.,11.)(0.,){/Sine}{0}
\FALabel(8.80315,8.72127)[tr]{$\mathrm{Z/\gamma}$}
\FAVert(3.5,10.){0}
\FAVert(11.,7.5){0}
\FAVert(16.,6.){0}
\FAVert(8.,11.){0}

\FADiagram{}
\FAProp(0.,15.)(8.,14.5)(0.,){/Straight}{1}
\FALabel(4.09669,15.817)[b]{$q$}
\FAProp(0.,5.)(8.,5.5)(0.,){/Cycles}{0}
\FALabel(4.09669,4.18302)[t]{$\mathrm{g}$}
\FAProp(20.,18.)(15.5,15.)(0.,){/Straight}{-1}
\FALabel(17.3702,17.3097)[br]{$l$}
\FAProp(20.,12.)(15.5,15.)(0.,){/Straight}{1}
\FALabel(17.3702,12.6903)[tr]{$l$}
\FAProp(20.,8.)(8.,5.5)(0.,){/Straight}{-1}
\FALabel(11.551,4.73525)[t]{$q$}
\FAProp(20.,2.)(8.,14.5)(0.,){/Sine}{0}
\FALabel(16.7997,3.80687)[tr]{$\gamma$}
\FAProp(8.,14.5)(8.,10.)(0.,){/Straight}{1}
\FALabel(6.93,12.25)[r]{$q$}
\FAProp(8.,5.5)(8.,10.)(0.,){/Straight}{-1}
\FALabel(6.93,7.75)[r]{$q$}
\FAProp(15.5,15.)(8.,10.)(0.,){/Sine}{0}
\FALabel(13.2061,14.5881)[br]{$\mathrm{Z/\gamma}$}
\FAVert(8.,14.5){0}
\FAVert(8.,5.5){0}
\FAVert(15.5,15.){0}
\FAVert(8.,10.){0}

\FADiagram{}
\FAProp(0.,15.)(8.,14.5)(0.,){/Straight}{1}
\FALabel(4.09669,15.817)[b]{$q$}
\FAProp(0.,5.)(8.,5.5)(0.,){/Cycles}{0}
\FALabel(4.09669,4.18302)[t]{$\mathrm{g}$}
\FAProp(20.,18.)(13.5,15.)(0.,){/Straight}{-1}
\FALabel(16.5805,17.4273)[br]{$l$}
\FAProp(20.,12.)(13.5,15.)(0.,){/Straight}{1}
\FALabel(16.5805,12.5727)[tr]{$l$}
\FAProp(20.,8.)(8.,5.5)(0.,){/Straight}{-1}
\FALabel(15.437,8.0267)[b]{$q$}
\FAProp(20.,2.)(8.,10.)(0.,){/Sine}{0}
\FALabel(16.0956,3.41321)[tr]{$\gamma$}
\FAProp(8.,14.5)(13.5,15.)(0.,){/Sine}{0}
\FALabel(10.6097,15.8136)[b]{$\mathrm{Z/\gamma}$}
\FAProp(8.,14.5)(8.,10.)(0.,){/Straight}{1}
\FALabel(6.93,12.25)[r]{$q$}
\FAProp(8.,5.5)(8.,10.)(0.,){/Straight}{-1}
\FALabel(6.93,7.75)[r]{$q$}
\FAVert(8.,14.5){0}
\FAVert(8.,5.5){0}
\FAVert(13.5,15.){0}
\FAVert(8.,10.){0}

\FADiagram{}
\FAProp(0.,15.)(8.,14.5)(0.,){/Straight}{1}
\FALabel(4.09669,15.817)[b]{$q$}
\FAProp(0.,5.)(8.,5.5)(0.,){/Cycles}{0}
\FALabel(4.09669,4.18302)[t]{$\mathrm{g}$}
\FAProp(20.,18.)(14.,15.)(0.,){/Straight}{-1}
\FALabel(16.7868,17.4064)[br]{$l$}
\FAProp(20.,12.)(14.,15.)(0.,){/Straight}{1}
\FALabel(16.7868,12.5936)[tr]{$l$}
\FAProp(20.,8.)(14.,5.)(0.,){/Straight}{-1}
\FALabel(16.7868,7.40636)[br]{$q$}
\FAProp(20.,2.)(14.,5.)(0.,){/Sine}{0}
\FALabel(16.7868,2.59364)[tr]{$\gamma$}
\FAProp(8.,14.5)(8.,5.5)(0.,){/Straight}{1}
\FALabel(6.93,10.)[r]{$q$}
\FAProp(8.,14.5)(14.,15.)(0.,){/Sine}{0}
\FALabel(10.8713,15.8146)[b]{$\mathrm{Z/\gamma}$}
\FAProp(8.,5.5)(14.,5.)(0.,){/Straight}{1}
\FALabel(10.8713,4.18535)[t]{$q$}
\FAVert(8.,14.5){0}
\FAVert(8.,5.5){0}
\FAVert(14.,15.){0}
\FAVert(14.,5.){0}

\FADiagram{}
\FAProp(0.,15.)(8.,14.5)(0.,){/Straight}{1}
\FALabel(4.09669,15.817)[b]{$q$}
\FAProp(0.,5.)(8.,5.5)(0.,){/Cycles}{0}
\FALabel(4.09669,4.18302)[t]{$\mathrm{g}$}
\FAProp(20.,18.)(16.,10.)(0.,){/Straight}{-1}
\FALabel(18.992,14.3042)[tl]{$l$}
\FAProp(20.,12.)(12.5,14.)(0.,){/Straight}{1}
\FALabel(15.0509,14.4149)[b]{$l$}
\FAProp(20.,8.)(8.,5.5)(0.,){/Straight}{-1}
\FALabel(11.551,4.73525)[t]{$q$}
\FAProp(20.,2.)(16.,10.)(0.,){/Sine}{0}
\FALabel(17.5842,4.83175)[tr]{$\gamma$}
\FAProp(8.,14.5)(8.,5.5)(0.,){/Straight}{1}
\FALabel(6.93,10.)[r]{$q$}
\FAProp(8.,14.5)(12.5,14.)(0.,){/Sine}{0}
\FALabel(10.4212,15.3105)[b]{$\mathrm{Z/\gamma}$}
\FAProp(16.,10.)(12.5,14.)(0.,){/Straight}{-1}
\FALabel(13.5635,11.4593)[tr]{$l$}
\FAVert(8.,14.5){0}
\FAVert(8.,5.5){0}
\FAVert(16.,10.){0}
\FAVert(12.5,14.){0}

\FADiagram{}
\FAProp(0.,15.)(8.,14.5)(0.,){/Straight}{1}
\FALabel(4.09669,15.817)[b]{$q$}
\FAProp(0.,5.)(8.,5.5)(0.,){/Cycles}{0}
\FALabel(4.09669,4.18302)[t]{$\mathrm{g}$}
\FAProp(20.,18.)(13.5,14.)(0.,){/Straight}{-1}
\FALabel(16.4176,16.8401)[br]{$l$}
\FAProp(20.,12.)(15.,9.5)(0.,){/Straight}{1}
\FALabel(18.3682,12.2436)[br]{$l$}
\FAProp(20.,8.)(8.,5.5)(0.,){/Straight}{-1}
\FALabel(11.4219,5.03777)[t]{$q$}
\FAProp(20.,2.)(15.,9.5)(0.,){/Sine}{0} 
\FALabel(17.944,3.59725)[tr]{$\gamma$}
\FAProp(8.,14.5)(8.,5.5)(0.,){/Straight}{1}
\FALabel(6.93,10.)[r]{$q$}
\FAProp(8.,14.5)(13.5,14.)(0.,){/Sine}{0}
\FALabel(10.8903,15.3136)[b]{$\mathrm{Z/\gamma}$}
\FAProp(13.5,14.)(15.,9.5)(0.,){/Straight}{-1}
\FALabel(13.2595,11.2598)[r]{$l$}
\FAVert(8.,14.5){0}
\FAVert(8.,5.5){0}
\FAVert(13.5,14.){0}
\FAVert(15.,9.5){0}

\end{feynartspicture}
\end{small}

\vspace*{-2em}
\end{center}
\mycaption{\label{fi:EW_real_udWg}
Real photonic bremsstrahlung corrections to the LO process \refeq{eq:proc2}.}
\efi

For muons in the final state it is experimentally possible to separate
collinear photons from the lepton, i.e.\ to observe so-called ``bare''
muons. On the theoretical side, this corresponds to the fact that the
lepton mass cuts off the collinear divergence in a physically
meaningful way. Hence, the corresponding collinear singularities show
up as logarithms of the small lepton (muon) mass since the KLN theorem
\cite{Kinoshita:1962ur} does not apply to non-collinear-safe
observables. Our treatment and the analytical extraction of these
collinear mass logarithms using the algorithm of
\citere{Dittmaier:2008md} has been explained in detail in
\citere{Denner:2009gj}. For $\PZ+\mathrm{jet}$ production, there are
of course two charged leptons in the final state which can emit
collinear photons. Apart from this straightforward generalization,
which simply leads to more dipole subtraction terms, the same
formalism applies.

As in all processes with jets in the final state, the inclusion of EW
corrections to $\PZ + \mathrm{jet}$ production asks for a precise event
definition in order to distinguish $\PZ\ + \mathrm{jet}$ from $\PZ +
\gamma$ production. We follow the strategy used for $\PW + \mathrm{jet}$
production which is detailed in
Refs.~\cite{Denner:2009gj,Denner:2010ia}: We exclude jets which
primarily consist of a hard photon (see Section \ref{se:numres}) and
capture the non-perturbative physics in the collinear quark--photon
splittings~\cite{Glover:1993xc} by means of the measured fragmentation
function~\cite{Buskulic:1995au}.  Note that our calculation also
provides an important step towards the inclusive prediction of dilepton
pairs at large transverse momentum up to
$\mathcal{O}(\alpha^3\alpha_\mathrm{s})$, i.e.\ dilepton pairs recoiling
against a jet or a photon, which would be free of the need to
distinguish a photon from a jet~\cite{Hollik:2007sq}. The only missing
ingredient is the combination of our results with the NLO QCD correction
for the dilepton+photon final
state~\cite{Campbell:2011bn,DeFlorian:2000sg}.  However, since
dilepton+photon production is suppressed with respect to the
dilepton+jet final state, the inclusive results will at most
differ at the level of one percent from the results presented in this work.\\

Concerning the real corrections in NLO QCD, there are no particular 
complications when all the contributing partonic channels are identified. 
An additional gluon leads to the processes 
\begin{eqnarray}
\label{eq:bremsproc4}
& \Pq_i \, \, \Pqbar_i & \to \Plp \Plm \, \Pg \, \Pg \, ,\\
\label{eq:bremsproc5}
& \Pq_i \, \, \Pg      & \to \Plp \Plm \, \Pq_i \, \Pg \, ,\\
\label{eq:bremsproc6}
& \Pqbar_i \, \, \Pg   & \to \Plp \Plm \, \Pqbar_i \, \Pg \, , \\
\label{eq:bremsproc7}
& \Pg\, \, \Pg   & \to \Plp \Plm \, \Pqbar_i \, \Pq_i \, .
\end{eqnarray}
Furthermore, the gluon present at LO may split into two quarks, 
inducing the processes
\begin{eqnarray}
\label{eq:bremsproc8}
& \Pq_i \, \, \Pqbar_i & \to \Plp \Plm \, \Pq_i \, \Pqbar_i \, ,\\
\label{eq:bremsproc9}
& q_i \, \, \Pqbar_i & \to \Plp \Plm \, \Pq_j \, \Pqbar_j 
\quad (\Pq_i\ne\Pq_j) \, ,\\
\label{eq:bremsproc10}
& q_i \, \, \Pqbar_j & \to \Plp \Plm \, \Pq_i \, \Pqbar_j 
\quad (\Pq_i\ne\Pq_j) \, ,\\
\label{eq:bremsproc11}
& \Pqbar_i \, \, \Pqbar_i & \to \Plp \Plm \, \Pqbar_i \, \Pqbar_i \, ,\\
\label{eq:bremsproc12}
& \Pqbar_i \, \, \Pqbar_j & \to \Plp \Plm \, \Pqbar_i \, \Pqbar_j 
\quad (\Pq_i\ne\Pq_j) \, ,\\
\label{eq:bremsproc13}
& \Pq_i \, \, \Pq_i & \to \Plp \Plm \, \Pq_i \, \Pq_i \, ,\\
\label{eq:bremsproc14}
& \Pq_i \, \, \Pq_j & \to \Plp \Plm \, \Pq_i \, \Pq_j
\quad (\Pq_i\ne\Pq_j) \, ,
\end{eqnarray}
where $q_i$ denotes again quarks of the five light flavours.  Note
that different Feynman diagrams contribute for $i=j$ and $i\ne j$, so
we list the corresponding partonic processes separately.  Taking this
difference into account, as well as the correct quantum numbers for
up- and down-type quarks, the remaining sums over flavour can again
efficiently be performed when convoluting the squared matrix elements
with PDFs. As in the EW case, we use the dipole subtraction method
\cite{Catani:1996vz} to extract the IR singularities analytically from
the numerical phase-space integration.  Absorbing all the collinear
singularities due to initial-state splittings into the relevant PDFs,
the remaining collinear and soft divergences cancel all the
divergences of the one-loop QCD corrections for processes
\refeq{eq:proc1}--\refeq{eq:proc3}. As explained already in
Section~\ref{se:virt}, we do not include the NLO QCD corrections to
the photon-induced processes.

For the six-fermion processes \refeq{eq:bremsproc8},
\refeq{eq:bremsproc11}, and \refeq{eq:bremsproc13} with identical
fermions, diagrams with gluon exchange can interfere with purely EW
diagrams at $\mathcal{O}(\alpha^3 \alpha_{\mathrm{s}})$.  The result
is non-singular in the collinear limits due to the restrictions from
colour flow, but in contrast to the other subprocesses with different
quark flavours it does not vanish. However, the effect of this
non-trivial interference contribution is expected to be
phenomenologically negligible, as shown for the similar process
$\PW+\mathrm{jet}$ production in \citere{Denner:2009gj}. Therefore, we
neglect these contributions in our numerical analysis.

\section{Numerical results}
\label{se:numres}

\subsection{Input parameters and setup}
\label{se:SMinput}

The relevant SM input parameters are
\begin{equation}\arraycolsep 2pt
\begin{array}[b]{lcllcllcl}
\GF & = & 1.16637 \times 10^{-5} \GeV^{-2}, \quad&
\alpha_{\mathrm{s}}(\MZ) &=& 0.1202 , 
&&&
\\
\MW^{\OS} & = & 80.398\GeV, &
\Gamma_\PW^{\OS} & = & 2.141\GeV, \\
\MZ^{\OS} & = & 91.1876\GeV, &
\Gamma_\PZ^{\OS} & = & 2.4952\GeV, &
M_\PH & = & 120\GeV, \\
m_\Pe & = & 0.510998910\MeV, &
m_\mu &=& 105.658367\MeV,\quad &
m_\Pt & = & 172.6\;\GeV,
\end{array}
\label{eq:SMpar}
\end{equation}
which essentially follow \citere{Amsler:2008zzb}. As stated before, the CKM
matrix only appears in loops and is 
set to unity, because its effect is negligible there.

Using the complex-mass scheme~\cite{Denner:2005fg}, we employ a fixed
width in the resonant W- and Z-boson propagators in contrast to the
approach used at LEP and Tevatron
to fit the W~and Z~resonances, where running
widths are taken. Therefore, we have to convert the ``on-shell'' (OS)
values of $M_V^{\OS}$ and $\Ga_V^{\OS}$ ($V=\PW,\PZ$), resulting
from LEP and Tevatron, to the ``pole values'' denoted by $M_V$ and $\Ga_V$. The
relation between the two sets of values is given by
\cite{Bardin:1988xt}
\beq\label{eq:m_ga_pole}
M_V = M_V^{\OS}/
\sqrt{1+(\Ga_V^{\OS}/M_V^{\OS})^2},
\qquad
\Ga_V = \Ga_V^{\OS}/
\sqrt{1+(\Ga_V^{\OS}/M_V^{\OS})^2},
\eeq
leading to
\beqar
\begin{array}[b]{r@{\,}l@{\qquad}r@{\,}l}
\MW &= 80.370\ldots\GeV, & \GW &= 2.1402\ldots\GeV, \\
\MZ &= 91.153\ldots\GeV,& \GZ &= 2.4943\ldots\GeV.
\label{eq:m_ga_pole_num}
\end{array}
\eeqar
We make use of these mass and width parameters in the numerics
discussed below, although the difference between using $M_V$ or
$M_V^{\OS}$ would be hardly visible.

As explained in \refse{se:setup}, we adopt the $\GF$ scheme, where the
electromagnetic coupling $\alpha$ is set to $\alpha_{\GF}$. 
In this scheme the electric-charge renormalization constant does not contain
logarithms of the light-fermion masses, in contrast to the $\alpha(0)$
scheme, so that the results become practically
independent of the light-quark masses.

We use the central MSTW2008NLO PDF set~\cite{Martin:2009iq} in its LHAPDF 
implementation~\cite{Whalley:2005nh} for our numerical results. This 
implies the value of $\alpha_{\mathrm{s}}(\MZ)$ stated in 
\refeq{eq:SMpar}. We use the $\alpha_{\mathrm{s}}$ running as
provided by the LHAPDF collaboration.  Only the 
photon-induced processes are evaluated with the MRSTQED2004 set of
PDFs~\cite{Martin:2004dh} as discussed in \refse{se:real}. Here, we
use the corresponding value of $\alpha_{\mathrm{s}}(\MZ) = 0.1190$.

The QCD and QED factorization scales as well as the renormalization
scale are always identified.  For low-$p_{\rT}$ jets, the scale of the
process is given by the invariant mass of the leptons which in turn
peaks around \MZ\ for resonant \PZ-boson production.  Hence, one
natural choice is the \PZ-boson mass, i.e.\ 
$\mu_\mathrm{R}=\mu_\mathrm{F}=\MZ$.  For high-$p_{\rT}$ jets, well
beyond the \PZ-mass scale, however, the relevant scale is certainly
larger, and the QCD emission from the initial state is best modelled
by the $p_{\rT}$ of the jet itself (see e.g.\ \citere{Bauer:2009km}).
To interpolate between the two regimes, we alternatively use
\begin{equation}
\label{eq:scale_choice}
\mu^{\mathrm{var}}=\mu^{\mathrm{var}}_\mathrm{R}=\mu^{\mathrm{var}}_\mathrm{F}=\sqrt{\MZ^2+(p_{\rT}^\mathrm{had})^2} \,  
\end{equation}
as a phase-space dependent scale, where $p_{\rT}^\mathrm{had}$ is
given by the $p_{\rT}$ of the summed four-momenta of all partons,
i.e.\ quarks and/or gluons, in the final state. At LO,
$p_{\rT}^\mathrm{had}$ is simply the $p_{\rT}$ of the one final-state
jet. We present numerical results for both scale choices.

\subsection{Phase-space cuts and event selection}
\label{se:cuts}

In order to define IR-safe observables for the process
$\Pp\Pp/\Pp\bar\Pp\to\PZ/\gamma^* + \mathrm{jet} \to \Pl^+\Pl^- +
\mathrm{jet} + \X$ we recombine final-state partons and photons to
pseudo-particles and impose a set of phase-space cuts as detailed in
the following subsections.

\subsubsection{Recombination}

To define the recombination procedure and the separation cuts, we use
the variables $R_{ij} = \sqrt{(y_{i}-y_{j})^2+\phi_{ij}^2}$, where
$y_{i}$ denotes the rapidity $y = \frac{1}{2} \ln [(E + p_\RL)/(E -
p_\RL)]$ of particle $i$ and $\phi_{ij}$ is the azimuthal angle in the
transverse plane between the particles $i$ and $j$. In the definition
of the rapidity, $E$ denotes the particle's energy and $p_{\RL}$ its
three-momentum along the beam axis. The recombination procedure,
where we simply add four-momenta to form a pseudo-particle, works as
follows:
\begin{enumerate}
\item For observables with bare muons we do not recombine photons and
  leptons. For inclusive observables, a photon and a lepton are
  recombined for $R_{\gamma l} < 0.1$. If both charged leptons in the
  final state are close to the photon we recombine it with the lepton
  leading to the smallest $R_{\gamma l}$.
\item A photon and a parton $a$ (quark or gluon) are recombined for
  $R_{\gamma a} < 0.5$. In this case, we use the energy fraction of
  the photon inside the jet, $z_{\gamma}=
  E_{\gamma}/(E_{\gamma}+E_a)$, to distinguish between
  $\mathrm{Z}+\mathrm{jet}$ and $\mathrm{Z}+\gamma$ production. If
  $z_{\gamma} > 0.7$, the event is regarded as a part of
  $\mathrm{Z}+\gamma$ production and rejected because it lacks any
  other hard jet at NLO. This event definition is not collinear safe
  and requires the use of quark-to-photon fragmentation functions to
  include the non-perturbative part of the quark--photon splitting as
  mentioned in \refse{se:real}.  Our results are not very sensitive to
  the specific choice of the cut on $z_{\gamma}$.
\item Two partons $a,b$ are recombined for $R_{ab} < 0.5$. For our
  simple final-state configurations, this procedure is equivalent to
  the Tevatron Run II $k_{\rT}$-algorithm \cite{Blazey:2000qt} for jet
  reconstruction with resolution parameter $D=0.5$.
\end{enumerate}

Technically, we perform a possible photon--lepton recombination before
the photon--parton recombination. This procedure is IR safe because
the triple-soft/collinear situation that a photon should have been
first recombined with a parton, but was erroneously first recombined
with a lepton, is excluded by our basic cuts.

\subsubsection{Basic cuts}
\label{se:basic_cuts}

After applying the recombination procedure of the previous section we define
$\PZ+\mathrm{jet}$ events by the following basic cuts:
\begin{enumerate}
\item A partonic object (after a possible recombination) is called a
  jet if its transverse momentum $p_{\rT}$ is larger than
  $p^{\mathrm{cut}}_{\rT,\mathrm{jet}} = 25\GeV$.  Events are required
  to include at least one jet.
\item
We demand two charged leptons with transverse momenta 
$p_{\rT,\Pl} > 25\GeV$.
\item
For the dilepton invariant mass we require
$M_{\Pl\Pl} > 50\GeV$ to cut events with nearly on-shell photons 
splitting into a collinear lepton pair.
\item
The events have to be central, i.e.\ the leptons and at least one jet have 
to be produced in the rapidity range $|y| < y_{\mathrm{max}} = 2.5$. 
\item The leptons have to be isolated, i.e.\ the event is discarded if
  the distance between one of the leptons and a jet
  $R_{l\mathrm{jet}}$ is smaller than 0.5. 
  The lepton--jet separation is also required for jets with $|y| >
  y_{\mathrm{max}}$.  
  
  Note in addition that it is important to exclude low-$p_{\rT}$
  partons from the lepton--jet separation procedure (guaranteed by
  step 1.), since otherwise observables would not be IR safe.
\end{enumerate} 

While the EW corrections differ for final-state electrons and
muons without photon recombination, the corrections become universal
in the presence of photon recombination, since the lepton-mass
logarithms cancel in this case, in accordance with the KLN theorem.
Numerical results are presented both
for photon recombination and for bare muons.

For certain observables, we apply a jet veto against a second hard jet.
To be specific, we veto any sub-leading jet with $p_{\rT} >
p_{\rT,\Pj_1}/2$, where $p_{\rT,\Pj_1}$ denotes the
$p_{\rT}$ of the ``leading'' jet, i.e.\ the one with maximal $p_{\rT}$.

\subsection{Results on cross sections and distributions}
\label{se:CSresults}

We consider the production of a lepton pair in association with a jet 
at the Tevatron, i.e.\ for a $\Pp\bar\Pp$~initial state with a 
centre-of-mass (CM) energy of $\sqrt{s}=1.96\TeV$, and at the LHC, 
i.e.\ for a \Pp\Pp~initial state. For the latter, we show results for 
$\sqrt{s}=7\TeV$, corresponding to the available energy in the years 
2010 to 2012, as well as $\sqrt{s}=14\TeV$, the ultimate energy reach 
of the LHC.

We present the LO cross section $\sigma_0$ and various types of
corrections $\de$, defined relative to the LO cross section by $\sigma
= \sigma_0\times\left(1+\de\right)$. Concerning the EW corrections, we
distinguish the cross section $\sigma^{\mu^+ \mu^-}_\mathrm{EW}$ for
bare muons and $\sigma^{\mathrm{rec}}_\mathrm{EW}$ for which a
lepton--photon recombination is employed as defined above.
Accordingly, the corresponding corrections are labelled $\delta^{\mu^+
  \mu^-}_\mathrm{EW}$ and $\delta^\mathrm{rec}_\mathrm{EW}$,
respectively. An additional label specifies which renormalization and
factorization scale is used. Either we use the fixed scale ($\mu=\MZ$)
or we determine the scale on an event-by-event basis by the
kinematical configuration of the final state (var), as specified in
\refeq{eq:scale_choice}.  For the EW corrections the difference is
small, since the LO and the NLO results depend on the renormalization
scale for $\alpha_{\mathrm{s}}$ and the QCD factorization scale in the
same way. However, for the QCD part a sensible scale choice can be
crucial for the stability of the perturbative series.  Accordingly,
the QCD corrections are labelled $\delta_\mathrm{QCD}^{\mu=\MZ}$ for a
fixed scale choice and $\delta_\mathrm{QCD}^\mathrm{var}$ for the
scale choice defined in \refeq{eq:scale_choice}.

As already observed for $\PW+\mathrm{jet}$ production in \citere{Denner:2009gj}
and shown below, the QCD corrections become larger and larger with
increasing $p_{\rT}$ of the leading jet.  The increase in the cross
section results from a new kinematical configuration which is
available for the $\PZ+2\,\mathrm{jets}$ final state.  The large
$p_{\rT}$ of the leading jet is not balanced by the leptons, as
required at LO, but by the second jet. Hence, we encounter the
production of 2 jets where one of the quark lines radiates a
relatively soft \PZ\ boson or off-shell photon. This part of the cross
section, which does not really correspond to a true NLO correction to
$\PZ+\mathrm{jet}$ production, can be separated by employing a veto
against a second hard jet in real-emission events. Hence, we present
NLO QCD corrections with a jet veto
($\delta^{\mu=\MZ}_\mathrm{QCD,veto}$,
$\delta^\mathrm{var}_\mathrm{QCD,veto}$) and without a jet veto
($\delta^{\mu=\MZ}_\mathrm{QCD}$, $\delta^\mathrm{var}_\mathrm{QCD}$).

Using a jet veto based on a fixed $p_{\rT}$ value for the second jet
is not well suited. It will either cut away relatively collinear
emission events in the high-$p_{\rT}$ tails of the leading-jet
distribution (leading to large negative corrections) or it has to be
chosen too large to be effective in the intermediate-$p_{\rT}$\ parts
of the distribution. Hence, building on our experience from
$\PW+\mathrm{jet}$ production, we apply the jet veto defined at the
end of \refse{se:basic_cuts}.  We have checked that this jet veto
indeed effectively removes events with back-to-back jet kinematics.

We also investigate the impact of the photon-induced tree-level
processes \refeq{eq:proc4} and \refeq{eq:proc5}. Since the LO
photon-induced cross section is a small effect, we show its relative
impact $\delta_{\gamma}$ with respect to the LO cross
section at $\mathcal{O}(\alpha^2\alpha_{\mathrm{s}})$ where initial
states with photons are not taken into account. The NLO QCD
corrections to these channels as well as the interference
contributions discussed at the end of \refse{se:real} are neglected in
the following.

\begin{table}
                                                                                                                                       $$ \begin{array}{c|rrrrrr}
                                                                   \multicolumn{7}{c}{\Pp\Pp \to \Plp \Plm\; \mathrm{jet} + X \;\mbox{at} \;\sqrt{s} =14 \TeV} \\
              \hline p_{\rT,\mathrm{jet}} / \GeV & 25-\infty \;\;\; & 50-\infty \;\;\; & 100-\infty \;\; & 200-\infty \;\; & 500-\infty \;\; & 1000-\infty \; \\ 
                                                                                                                                                     \hline\hline
\si_{0}^{\mu = \MZ}/\fba                          \; & \; 123491(7)       \; & \; 44603(2)        \; & \; 11364.4(5)      \; & \; 1813.26(8)      \; & \; 64.120(2)       \; & \; 2.11859(6)      \\ 
\si_{0}^{\mathrm{var}}/\fba                       \; & \; 122024(7)       \; & \; 43254(2)        \; & \; 10445.0(4)      \; & \; 1475.76(6)      \; & \; 38.648(1)       \; & \; 0.90847(3)      \\ 
   \hline \hline                                                                                                         
\de_{\EW}^{\mu^+\mu^-\,,\mathrm{var}}/\%              \; & \; -4.2\phz        \; & \; -4.5\phz        \; & \; -5.1\phz        \; & \; -8.5\phz        \; & \; -17.4(1)        \; & \; -27.0(1)        \\ 
\de_{\EW}^{\mathrm{rec}\,,\mathrm{var}}/\%            \; & \; -2.8\phz        \; & \; -3.2\phz        \; & \; -4.2\phz        \; & \; -7.8\phz        \; & \; -16.7(1)        \; & \; -26.3(1)        \\ 
   \hline \hline                                                                                                         
\de_{\QCD}^{\mu = \MZ}/\%                             \; & \; 35.8(1)         \; & \; 48.7(1)         \; & \; 63.9(1)         \; & \; 86.9(1)         \; & \; 142.6(1)        \; & \; 210.5(1)        \\ 
\de_{\QCD}^{\mathrm{var}}/\%                          \; & \; 35.9(1)         \; & \; 50.1(1)         \; & \; 70.0(1)         \; & \; 107.0(1)        \; & \; 217.3(1)        \; & \; 403.8(1)        \\ 
  \hline                                                                                                                 
\de_{\QCD,\veto}^{\mu = \MZ}/\%                       \; & \; 13.1(1)         \; & \; 9.8(1)          \; & \; 14.1(1)         \; & \; 13.4(1)         \; & \; -2.4(1)         \; & \; -29.1(1)        \\ 
\de_{\QCD,\veto}^{\mathrm{var}}/\%                    \; & \; 14.1(1)         \; & \; 12.6(1)         \; & \; 21.9(1)         \; & \; 32.3(1)         \; & \; 44.8(1)         \; & \; 54.2(1)         \\ 
   \hline \hline                                                                                                         
\de_{\ga}^{\mathrm{var}}/\%                     \; & \; 0.1\phz         \; & \; 0.2\phz         \; & \; 0.2\phz         \; & \; 0.4\phz         \; & \; 0.6\phz         \; & \; 1.0\phz         \\ 
   \hline \hline                                                                                                         
\si_{\mathrm{full, veto}}^{\mu^+\mu^-\,,\mathrm{var}}/\fba \; & \; 134266(49)      \; & \; 46852(20)       \; & \; 12223(4)        \; & \; 1832.5(8)       \; & \; 49.45(2)        \; & \; 1.1649(7)       \\ 
 \end{array} $$

\mycaption{\label{ta:ptj_LHC} Integrated cross sections for different
  cuts on the $p_{\rT}$ of the leading jet (jet with highest $p_{\rT}$) at the LHC with 
  $\sqrt{s}=14\TeV$. We show the LO
  results both for a variable and for a constant scale. The relative EW
  corrections $\de_{\EW}$ are given with and without lepton--photon
  recombination. The QCD corrections $\de_{\QCD}$ are presented for 
  a fixed as well as a for variable scale and with or without employing
  a veto on a second hard jet. The EW corrections and the corrections due to photon-induced 
  processes, $\de_{\ga}$, are presented for the variable
  scale. Finally, we show the full NLO cross section
  $\si_{\mathrm{full},\mathrm{veto}}^{\mu^+\mu^-,\mathrm{var}}$.
  The error from the Monte Carlo integration for the last digit(s)
  is given in parenthesis as far as significant. See text for details.
  }
\end{table}
\begin{table}
                                                                                                                                       $$ \begin{array}{c|rrrrrr}
                                                                    \multicolumn{7}{c}{\Pp\Pp \to \Plp \Plm\; \mathrm{jet} + X \;\mbox{at} \;\sqrt{s} =7 \TeV} \\
              \hline p_{\rT,\mathrm{jet}} / \GeV & 25-\infty \;\;\; & 50-\infty \;\;\; & 100-\infty \;\; & 200-\infty \;\; & 500-\infty \;\; & 1000-\infty \; \\ 
                                                                                                                                                     \hline\hline
\si_{0}^{\mu = \MZ}/\fba                          \; & \; 53029(3)        \; & \; 17736.0(5)      \; & \; 3939.51(9)      \; & \; 471.85(1)       \; & \; 7.4538(2)       \; & \; 0.06464(6)      \\ 
\si_{0}^{\mathrm{var}}/\fba                       \; & \; 51949(3)        \; & \; 16881.3(5)      \; & \; 3482.85(6)      \; & \; 357.071(7)      \; & \; 3.90038(8)      \; & \; 0.02139(2)      \\ 
   \hline \hline                                                                                                         
\de_{\EW}^{\mu^+\mu^-\,,\mathrm{var}}/\%              \; & \; -4.2\phz        \; & \; -4.4\phz        \; & \; -4.9\phz        \; & \; -8.0\phz        \; & \; -16.6(1)        \; & \; -26.0(1)        \\ 
\de_{\EW}^{\mathrm{rec}\,,\mathrm{var}}/\%            \; & \; -2.7\phz        \; & \; -3.1\phz        \; & \; -4.0\phz        \; & \; -7.3\phz        \; & \; -15.9(1)        \; & \; -25.2(1)        \\ 
   \hline \hline                                                                                                         
\de_{\QCD}^{\mu = \MZ}/\%                             \; & \; 35.1(1)         \; & \; 43.3(1)         \; & \; 52.2(1)         \; & \; 68.9(1)         \; & \; 119.2(1)        \; & \; 191.6(1)        \\ 
\de_{\QCD}^{\mathrm{var}}/\%                          \; & \; 36.3(1)         \; & \; 47.0(1)         \; & \; 63.9(1)         \; & \; 100.9(1)        \; & \; 229.7(1)        \; & \; 505.0(2)        \\ 
  \hline                                                                                                                 
\de_{\QCD,\veto}^{\mu = \MZ}/\%                       \; & \; 15.4(1)         \; & \; 8.7(1)          \; & \; 8.9(1)          \; & \; 4.0(1)          \; & \; -15.3(1)        \; & \; -44.9(1)        \\ 
\de_{\QCD,\veto}^{\mathrm{var}}/\%                    \; & \; 17.2(1)         \; & \; 13.4(1)         \; & \; 20.7(1)         \; & \; 30.9(1)         \; & \; 48.2(1)         \; & \; 69.4(1)         \\ 
   \hline \hline                                                                                                         
\de_{\ga}^{\mathrm{var}}/\%                     \; & \; 0.2\phz         \; & \; 0.2\phz         \; & \; 0.3\phz         \; & \; 0.4\phz         \; & \; 0.8\phz         \; & \; 1.9\phz         \\ 
   \hline \hline                                                                                                         
\si_{\mathrm{full, veto}}^{\mu^+\mu^-\,,\mathrm{var}} / \fba \; & \; 58823(23)       \; & \; 18438(8)        \; & \; 4045(2)         \; & \; 440.2(2)        \; & \; 5.168(2)        \; & \; 0.03106(2)      \\ 
 \end{array} $$

\mycaption{\label{ta:ptj_LHC7T} Integrated cross sections for different
  cuts on the $p_{\rT}$ of the leading jet at the LHC with $\sqrt{s}=7\TeV$. 
  See caption of \refta{ta:ptj_LHC} and text for details. 
  }
\end{table}

In \reftas{ta:ptj_LHC}--\ref{ta:mll_TEV} we show the LO integrated
cross sections, the corresponding relative corrections introduced
above, and the NLO cross section
$\si_{\mathrm{full},\mathrm{veto}}^{\mu^+\mu^-,\mathrm{var}}$ including the EW
corrections for bare muons, the photon-induced processes, and the QCD 
corrections with the jet veto for the variable scale choice
for different cuts on the transverse momentum of the leading jet and
the dilepton invariant mass.  All other cuts and the corresponding
event selection follow our default choice as introduced in
\refse{se:cuts}.  In \reffis{fi:ptj_all}--\ref{fi:yZ_all} we show
for various observables the LO distribution and the distribution
including the full set of corrections, \ie EW corrections
$\delta_{\EW}$, the contribution of the photon-induced processes
$\de_{\ga}$, and the QCD corrections $\de_{\QCD}$. The
various contributions to the corrections are also shown separately
relative to the LO. All results are discussed in detail in the
following subsections.

\subsubsection{Transverse momentum of the leading jet}

\reftas{ta:ptj_LHC}--\ref{ta:ptj_TEV} show the LO predictions and the
above corrections for different cuts on the $p_{\rT}$ of the leading
jet $p_{\rT,\mathrm{jet}}$. All integrated cross sections and, hence,
the corrections are dominated by events close to the lowest accepted
$p_{\rT,\mathrm{jet}}$, as can be seen by the rapid decrease of the
integrated cross section when increasing the $p_{\rT,\mathrm{jet}}$
cut.

For the most inclusive cross sections (left columns in the tables) 
the EW corrections are at the percent level and negative. 
With increasing $p_{\rT,\mathrm{jet}}$, the relevant CM energies
rise, and the well-known Sudakov logarithms in the virtual EW
corrections start to dominate the total corrections as expected. For
$p_{\rT,\mathrm{jet}} \sim 1000\GeV$, the EW corrections are at the level of
$-25\%$. This behaviour is generic and also present in all other observables
where the cross section is dominated by events with high CM energies.

It is evident from \reftas{ta:ptj_LHC} and \ref{ta:ptj_LHC7T} that the
CM energy of the LHC plays a minor role for the EW corrections, in
particular for less restrictive cuts. Since the transverse momentum of
the leading jet depends only indirectly on the treatment of the
lepton--photon system the corrections with and without recombination
only slightly differ. The results for the corresponding differential
distribution are shown in \reffi{fi:ptj_all} for the LHC with
$\sqrt{s}=14\TeV$ and the Tevatron.

\begin{table}
                                                                                                                                       $$ \begin{array}{c|rrrrrr}
                                                             \multicolumn{7}{c}{\Pp\bar\Pp \to \Plp \Plm\; \mathrm{jet} + X \;\mbox{at} \;\sqrt{s} =1.96 \TeV} \\
            \hline p_{\rT,\mathrm{jet}} / \GeV & 25-\infty \;\;\; & 50-\infty \;\;\; & 75-\infty \;\;\; & 100-\infty \;\; & 200-\infty \;\; & 300-\infty \;\; \\ 
                                                                                                                                                     \hline\hline
\si_{0}^{\mu = \MZ}/\fba                          \; & \; 9648.3(3)       \; & \; 2440.3(1)       \; & \; 840.01(5)       \; & \; 340.51(2)       \; & \; 17.679(1)       \; & \; 1.3935(1)       \\ 
\si_{0}^{\mathrm{var}}/\fba                       \; & \; 9365.0(4)       \; & \; 2268.2(1)       \; & \; 741.47(5)       \; & \; 284.88(2)       \; & \; 12.0551(9)      \; & \; 0.79178(7)      \\ 
   \hline \hline                                                                                                         
\de_{\EW}^{\mu^+\mu^-\,,\mathrm{var}}/\%              \; & \; -4.1\phz        \; & \; -4.2\phz        \; & \; -4.0\phz        \; & \; -4.2\phz        \; & \; -6.3\phz        \; & \; -8.4\phz        \\ 
\de_{\EW}^{\mathrm{rec}\,,\mathrm{var}}/\%            \; & \; -2.6\phz        \; & \; -2.8\phz        \; & \; -2.9\phz        \; & \; -3.3\phz        \; & \; -5.5\phz        \; & \; -7.7\phz        \\ 
   \hline \hline                                                                                                         
\de_{\QCD}^{\mu = \MZ}/\%                             \; & \; 30.7(1)         \; & \; 25.7(1)         \; & \; 20.3(1)         \; & \; 14.2(1)         \; & \; -8.7(1)         \; & \; -31.0(1)        \\ 
\de_{\QCD}^{\mathrm{var}}/\%                          \; & \; 32.9(1)         \; & \; 31.9(1)         \; & \; 31.9(1)         \; & \; 31.8(1)         \; & \; 33.4(1)         \; & \; 33.8(1)         \\ 
  \hline                                                                                                                 
\de_{\QCD,\veto}^{\mu = \MZ}/\%                       \; & \; 19.8(1)         \; & \; 5.8(1)          \; & \; 0.4(1)          \; & \; -5.6(1)         \; & \; -29.0(1)        \; & \; -50.6(1)        \\ 
\de_{\QCD,\veto}^{\mathrm{var}}/\%                    \; & \; 22.5(1)         \; & \; 12.4(1)         \; & \; 12.3(1)         \; & \; 11.4(1)         \; & \; 9.1(1)          \; & \; 7.1(1)          \\ 
   \hline \hline                                                                                                         
\de_{\ga}^{\mathrm{var}}/\%                     \; & \; 0.2\phz         \; & \; 0.2\phz         \; & \; 0.3\phz         \; & \; 0.3\phz         \; & \; 0.3\phz         \; & \; 0.3\phz         \\ 
   \hline \hline                                                                                                         
\si_{\mathrm{full, veto}}^{\mu^+\mu^-\,,\mathrm{var}} / \fba \; & \; 11102(6)        \; & \; 2458(2)         \; & \; 804(1)          \; & \; 306.2(3)        \; & \; 12.43(1)        \; & \; 0.784(1)        \\ 
 \end{array} $$

\mycaption{\label{ta:ptj_TEV} Integrated cross sections for different
  cuts on the $p_{\rT}$ of the leading jet at the Tevatron. See caption
  of 
  \refta{ta:ptj_LHC} and text for details.}
\end{table}

\bfi     
\bce
\includegraphics[width=15.7cm]{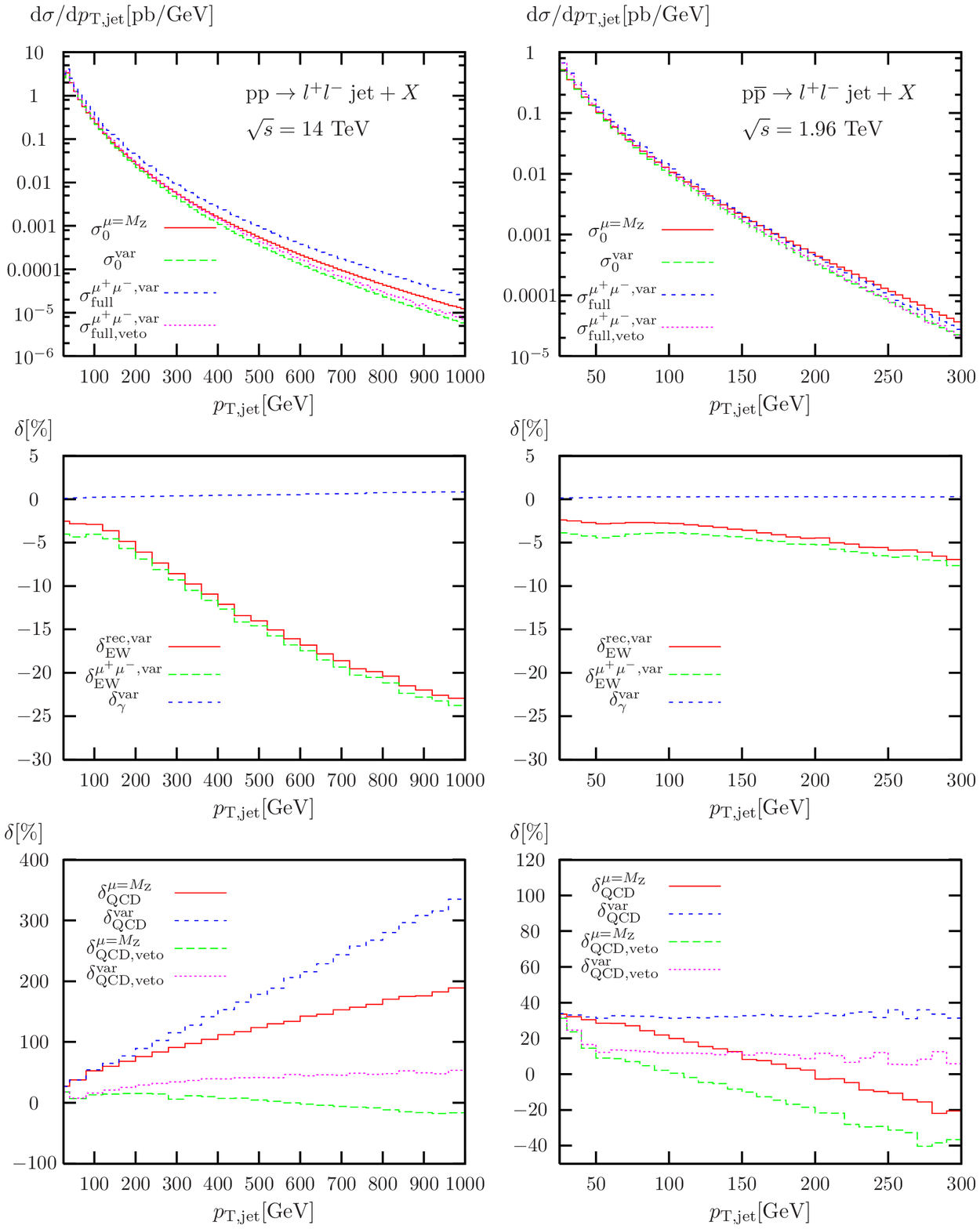} 
\ece
\mycaption{\label{fi:ptj_all} LO and fully corrected distribution (top),
  corresponding relative EW and photon-induced corrections (middle), and 
  relative QCD corrections (bottom) for the transverse momentum
  of the leading jet at the LHC (left) and the Tevatron (right). 
  See text for details.
  }
\efi 

The transverse-momentum distribution of the jet has been calculated
using the approximation of a stable, on-shell \PZ\ boson in
\citeres{Kuhn:2004em,Kuhn:2005az}. Of course, in contrast to our
calculation, in on-shell $\PZ+\mathrm{jet}$ production it is not
possible to apply various event-selection cuts to the leptonic final
state, because the degrees of freedom related to the decaying \PZ\ 
boson are implicitly integrated out. Nevertheless, the relative EW
corrections at high momentum transfer are dominated by Sudakov
logarithms of the form $\ln^2(\hat{s}/M_{\PZ}^2)$ that give rise to
large process-independent contributions which factorize from the LO
cross section. Therefore, they are expected to result in a similar
behaviour for both the on- and off-shell corrections.  Comparing our
results for the leading-jet $p_{\rT,\mathrm{jet}}$ in
\reffi{fi:ptj_all} with Figure\ 5 in \citere{Kuhn:2005az}, we find
agreement within $1{-}2\%$ for $p_{\rT,\mathrm{jet}}>200\GeV$. Only at
smaller $p_{\rT,\mathrm{jet}}$, the details of the event definition
start to be more relevant and the on-shell results deviate as
expected. However, the good agreement between the two calculations is
still remarkable, having in mind that in \citere{Kuhn:2005az} both the
QED corrections have been completely neglected and a different
renormalization scheme ($\overline{\mathrm{MS}}$) has been adopted.

The contribution $\delta_\gamma$ from the photon-induced processes are
small and only reach up to a few percent for large cut values where the
EW and QCD corrections to the dominating tree processes are by far
larger. Hence, we can safely neglect the corresponding NLO
QCD corrections which are formally of the same order as the EW
corrections to the partonic processes without a photon in the initial
state.

The qualitative features of the corrections at the Tevatron are very
similar to those at the LHC. Of course, at the Tevatron the high-energy
(Sudakov) regime is not as accessible as at the LHC, but the onset of the
Sudakov dominance is nevertheless visible as can be seen in
\refta{ta:ptj_TEV} and \reffi{fi:ptj_all}. We have
adapted the range for the different integrated cross sections to the
kinematic reach of the Tevatron.

Turning to the NLO QCD results at the LHC, we observe exactly the same
qualitative results found for $\PW+\mathrm{jet}$ production in
\citere{Denner:2009gj}.  As discussed above, the differential cross
section for large $p_{\rT,\mathrm{jet}}$, as shown in
\reffi{fi:ptj_all}, contains large contributions from a completely
different class of events for which two jets recoil against each
other. Hence, the corrections are huge. The correction
$\delta_\mathrm{QCD}^{\mu=\MZ}$ is smaller than
$\delta_\mathrm{QCD}^{\mathrm{var}}$ because it is defined relative to a
larger LO cross section. In absolute size, however, the NLO
corrections are similar. Using the jet veto proposed at the end of
\refse{se:basic_cuts}, the corrections are reduced, and
$\delta_\mathrm{QCD}^{\mathrm{var}}$ rises only to the 50\% level for
large cut values. The fixed scale choice accidentally leads to even
smaller corrections $\delta_\mathrm{QCD}^{\mu=\MZ}$.  As expected, the
large discrepancy of the LO results for the two scale choices is largely
removed by including the NLO corrections. However, varying the exact
definition of the jet veto, the variable scale turns out to be more
robust.

For the smaller CM energy $\sqrt{s}=7\TeV$ the same qualitative behaviour 
is found (see \refta{ta:ptj_LHC7T} for the integrated results with varying
cuts).
At the Tevatron, the jet veto is not as important as at the LHC
because of its kinematical limitations.
On the other hand, as expected, the fixed scale choice
leads to more and more negative corrections
with increasing $p_{\rT,\mathrm{jet}}$, in particular when employing
a jet veto. Using the variable scale stable results are obtained.

\subsubsection{Dilepton invariant mass}
\label{se:dileptonmass} 

\begin{table}
                                                                                                                                       $$ \begin{array}{c|rrrrrr}
                                                                   \multicolumn{7}{c}{\Pp\Pp \to \Plp \Plm\; \mathrm{jet} + X \;\mbox{at} \;\sqrt{s} =14 \TeV} \\
                    \hline M_{\Pl\Pl} / \GeV & 50-\infty \;\;\; & 100-\infty \;\; & 200-\infty \;\; & 500-\infty \;\; & 1000-\infty \; & 2000-\infty \;\; \\ 
                                                                                                                                                     \hline\hline
\si_{0}^{\mu = \MZ}/\fba                          \; & \; 123491(7)       \; & \; 7696.9(8)       \; & \; 628.47(6)       \; & \; 49.380(6)       \; & \; 5.1124(6)       \; & \; 0.27096(3)      \\ 
\si_{0}^{\mathrm{var}}/\fba                       \; & \; 122024(7)       \; & \; 7558.2(8)       \; & \; 602.45(5)       \; & \; 45.750(5)       \; & \; 4.5919(6)       \; & \; 0.23433(3)      \\ 
\si_{0}^{\mathrm{var},M_{ll}}/\fba                \; & \; 121888(7)       \; & \; 7419.8(8)       \; & \; 539.74(5)       \; & \; 34.102(4)       \; & \; 2.7958(4)       \; & \; 0.10831(1)      \\ 
   \hline \hline                                                                                                         
\de_{\EW}^{\mu^+\mu^-\,,\mathrm{var}}/\%              \; & \; -4.2\phz        \; & \; -9.3(1)         \; & \; -5.7\phz        \; & \; -9.5\phz        \; & \; -15.1(1)        \; & \; -23.8(1)        \\ 
\de_{\EW}^{\mathrm{rec}\,,\mathrm{var}}/\%            \; & \; -2.8\phz        \; & \; -5.2\phz        \; & \; -3.0\phz        \; & \; -5.8\phz        \; & \; -10.3(1)        \; & \; -17.1(1)        \\ 
   \hline \hline                                                                                                         
\de_{\QCD}^{\mu = \MZ}/\%                             \; & \; 35.8(1)         \; & \; 28.9(1)         \; & \; 12.0(1)         \; & \; -11.3(1)        \; & \; -34.4(1)        \; & \; -62.7(1)        \\ 
\de_{\QCD}^{\mathrm{var}}/\%                          \; & \; 35.9(1)         \; & \; 29.7(1)         \; & \; 14.7(1)         \; & \; -5.7(1)         \; & \; -25.8(1)        \; & \; -50.8(1)        \\ 
\de_{\QCD}^{\mathrm{var},M_{ll}}/\%                   \; & \; 36.1(1)         \; & \; 30.8(1)         \; & \; 24.7(1)         \; & \; 23.6(1)         \; & \; 25.9(3)         \; & \; 31.4(3)         \\ 
  \hline                                                                                                                 
\de_{\QCD,\veto}^{\mu = \MZ}/\%                       \; & \; 13.1(1)         \; & \; 6.8(1)          \; & \; -9.5(1)         \; & \; -32.9(1)        \; & \; -56.3(1)        \; & \; -85.3(1)        \\ 
\de_{\QCD,\veto}^{\mathrm{var}}/\%                    \; & \; 14.1(1)         \; & \; 8.7(1)          \; & \; -5.4(1)         \; & \; -25.4(1)        \; & \; -46.0(1)        \; & \; -71.3(1)        \\ 
\de_{\QCD,\veto}^{\mathrm{var},M_{ll}}/\%             \; & \; 14.3(1)         \; & \; 10.7(2)         \; & \; 7.4(1)          \; & \; 8.0(1)          \; & \; 10.9(3)         \; & \; 16.7(3)         \\ 
   \hline \hline                                                                                                         
\de_{\ga}^{\mathrm{var}}/\%                     \; & \; 0.1\phz         \; & \; 0.9\phz         \; & \; 2.7\phz         \; & \; 2.9\phz         \; & \; 2.6\phz         \; & \; 2.3\phz         \\ 
   \hline \hline                                                                                                         
\si_{\mathrm{full, veto}}^{\mu^+\mu^-\,,\mathrm{var}} / \fba \; & \; 134266(49)      \; & \; 7580(9)         \; & \; 551.9(4)        \; & \; 31.10(5)        \; & \; 1.906(4)        \; & \; 0.0167(3)       \\ 
 \end{array} $$

\mycaption{\label{ta:mll_LHC} Integrated cross sections for different
  cuts on the dilepton invariant mass $M_{\Pl\Pl}$ at the LHC with 
  $\sqrt{s}=14\TeV$. 
  In addition to the cross sections and corrections given in 
  \reftas{ta:ptj_LHC}--\ref{ta:ptj_TEV} we also show results for 
  an alternative variable scale choice introduced at the end of \refse{se:dileptonmass}.
  See text for details.}
\end{table}

\begin{table}
                                                                                                                                       $$ \begin{array}{c|rrrrrr}
                                                                    \multicolumn{7}{c}{\Pp\Pp \to \Plp \Plm\; \mathrm{jet} + X \;\mbox{at} \;\sqrt{s} =7 \TeV} \\
                    \hline M_{\Pl\Pl} / \GeV & 50-\infty \;\;\; & 100-\infty \;\; & 200-\infty \;\; & 500-\infty \;\; & 1000-\infty \; & 2000-\infty \;\; \\ 
                                                                                                                                                     \hline\hline
\si_{0}^{\mu = \MZ}/\fba                          \; & \; 53029(3)        \; & \; 3259.1(4)       \; & \; 242.06(3)       \; & \; 13.300(2)       \; & \; 0.7276(1)       \; & \; 0.009264(9)     \\ 
\si_{0}^{\mathrm{var}}/\fba                       \; & \; 51949(3)        \; & \; 3171.4(4)       \; & \; 230.25(2)       \; & \; 12.283(2)       \; & \; 0.65366(9)      \; & \; 0.008059(8)     \\ 
\si_{0}^{\mathrm{var},M_{ll}}/\fba                \; & \; 51865(3)        \; & \; 3064.5(4)       \; & \; 194.99(2)       \; & \; 8.235(1)        \; & \; 0.33869(5)      \; & \; 0.002829(1)    \\ 
   \hline \hline                                                                                                         
\de_{\EW}^{\mu^+\mu^-\,,\mathrm{var}}/\%              \; & \; -4.2\phz        \; & \; -9.3\phz        \; & \; -5.8\phz        \; & \; -10.2\phz       \; & \; -16.7(1)        \; & \; -27.7(1)        \\ 
\de_{\EW}^{\mathrm{rec}\,,\mathrm{var}}/\%            \; & \; -2.7\phz        \; & \; -5.2\phz        \; & \; -3.0\phz        \; & \; -6.1\phz        \; & \; -11.0(1)        \; & \; -19.3(1)        \\ 
   \hline \hline                                                                                                         
\de_{\QCD}^{\mu = \MZ}/\%                             \; & \; 35.1(1)         \; & \; 28.3(1)         \; & \; 11.0(1)         \; & \; -11.8(1)        \; & \; -34.0(1)        \; & \; -63.5(7)        \\ 
\de_{\QCD}^{\mathrm{var}}/\%                          \; & \; 36.3(1)         \; & \; 30.4(1)         \; & \; 14.8(1)         \; & \; -5.3(1)         \; & \; -25.0(1)        \; & \; -50.4(2)        \\ 
\de_{\QCD}^{\mathrm{var},M_{ll}}/\%                   \; & \; 36.5(1)         \; & \; 32.6(1)         \; & \; 29.4(1)         \; & \; 32.4(2)         \; & \; 38.9(3)         \; & \; 53.0(3)         \\ 
  \hline                                                                                                                 
\de_{\QCD,\veto}^{\mu = \MZ}/\%                       \; & \; 15.4(1)         \; & \; 9.3(1)          \; & \; -6.8(1)         \; & \; -30.0(1)        \; & \; -52.7(1)        \; & \; -81.3(2)        \\ 
\de_{\QCD,\veto}^{\mathrm{var}}/\%                    \; & \; 17.2(1)         \; & \; 12.0(1)         \; & \; -2.1(1)         \; & \; -22.2(1)        \; & \; -42.3(1)        \; & \; -67.1(1)        \\ 
\de_{\QCD,\veto}^{\mathrm{var},M_{ll}}/\%             \; & \; 17.6(1)         \; & \; 15.3(1)         \; & \; 14.7(1)         \; & \; 19.1(2)         \; & \; 26.1(3)         \; & \; 41.0(3)         \\ 
   \hline \hline                                                                                                         
\de_{\ga}^{\mathrm{var}}/\%                     \; & \; 0.2\phz         \; & \; 1.0\phz         \; & \; 2.9\phz         \; & \; 3.0\phz         \; & \; 2.8\phz         \; & \; 3.0\phz         \\ 
   \hline \hline                                                                                                         
\si_{\mathrm{full, veto}}^{\mu^+\mu^-\,,\mathrm{var}} / \fba \; & \; 58823(23)       \; & \; 3290(4)         \; & \; 218.7(1)        \; & \; 8.67(1)         \; & \; 0.2861(6)       \; & \; 0.00065(1)      \\ 
 \end{array} $$

\mycaption{\label{ta:mll_LHC7T} Integrated cross sections for different
  cuts on the dilepton invariant mass $M_{\Pl\Pl}$ at the LHC with 
  $\sqrt{s}=7\TeV$. See caption of \refta{ta:mll_LHC} and text for details.}
\end{table}

\begin{table}
                                                                                                                                       $$ \begin{array}{c|rrrrrr}
                                                             \multicolumn{7}{c}{\Pp\bar\Pp \to \Plp \Plm\; \mathrm{jet} + X \;\mbox{at} \;\sqrt{s} =1.96 \TeV} \\
                    \hline M_{\Pl\Pl} / \GeV & 50-\infty \;\;\; & 100-\infty \;\; & 150-\infty \;\; & 200-\infty \;\; & 400-\infty \;\; & 600-\infty \;\; \\ 
                                                                                                                                                     \hline\hline
\si_{0}^{\mu = \MZ}/\fba                          \; & \; 9648.3(3)       \; & \; 636.46(9)       \; & \; 121.20(1)       \; & \; 51.243(6)       \; & \; 3.9644(5)       \; & \; 0.41195(5)      \\ 
\si_{0}^{\mathrm{var}}/\fba                       \; & \; 9365.0(4)       \; & \; 614.55(9)       \; & \; 115.95(1)       \; & \; 48.770(5)       \; & \; 3.7177(4)       \; & \; 0.38255(5)      \\ 
\si_{0}^{\mathrm{var},M_{ll}}/\fba                \; & \; 9335.9(3)       \; & \; 574.15(8)       \; & \; 95.98(1)        \; & \; 37.349(4)       \; & \; 2.2362(3)       \; & \; 0.18691(2)      \\ 
   \hline \hline                                                                                                         
\de_{\EW}^{\mu^+\mu^-\,,\mathrm{var}}/\%              \; & \; -4.1\phz        \; & \; -9.0(1)         \; & \; -5.8\phz        \; & \; -6.5\phz        \; & \; -11.1\phz       \; & \; -15.7(1)        \\ 
\de_{\EW}^{\mathrm{rec}\,,\mathrm{var}}/\%            \; & \; -2.6\phz        \; & \; -5.2\phz        \; & \; -3.1\phz        \; & \; -3.5\phz        \; & \; -6.6\phz        \; & \; -9.5(1)         \\ 
   \hline \hline                                                                                                         
\de_{\QCD}^{\mu = \MZ}/\%                             \; & \; 30.7(1)         \; & \; 24.3(1)         \; & \; 15.3(1)         \; & \; 10.7(1)         \; & \; -0.3(1)         \; & \; -8.0(2)         \\ 
\de_{\QCD}^{\mathrm{var}}/\%                          \; & \; 32.9(1)         \; & \; 26.7(2)         \; & \; 18.5(1)         \; & \; 14.7(1)         \; & \; 4.9(1)          \; & \; -2.0(1)         \\ 
\de_{\QCD}^{\mathrm{var},M_{ll}}/\%                   \; & \; 33.3(1)         \; & \; 32.3(1)         \; & \; 34.6(1)         \; & \; 37.5(2)         \; & \; 49.1(2)         \; & \; 60.8(2)         \\ 
  \hline                                                                                                                 
\de_{\QCD,\veto}^{\mu = \MZ}/\%                       \; & \; 19.8(1)         \; & \; 13.8(2)         \; & \; 5.7(2)          \; & \; 1.1(1)          \; & \; -9.9(1)         \; & \; -17.4(2)        \\ 
\de_{\QCD,\veto}^{\mathrm{var}}/\%                    \; & \; 22.5(1)         \; & \; 16.7(2)         \; & \; 9.6(2)          \; & \; 5.7(1)          \; & \; -4.3(1)         \; & \; -10.7(1)        \\ 
\de_{\QCD,\veto}^{\mathrm{var},M_{ll}}/\%             \; & \; 22.8(1)         \; & \; 23.0(1)         \; & \; 26.2(1)         \; & \; 29.3(1)         \; & \; 41.6(2)         \; & \; 53.8(2)         \\ 
   \hline \hline                                                                                                         
\de_{\ga}^{\mathrm{var}}/\%                     \; & \; 0.2\phz         \; & \; 0.7\phz         \; & \; 1.2\phz         \; & \; 1.0\phz         \; & \; 0.5\phz         \; & \; 0.3\phz         \\ 
   \hline \hline                                                                                                         
\si_{\mathrm{full, veto}}^{\mu^+\mu^-\,,\mathrm{var}} / \fba \; & \; 11102(6)        \; & \; 665(1)          \; & \; 121.7(2)        \; & \; 48.93(5)        \; & \; 3.162(2)        \; & \; 0.2825(4)       \\ 
 \end{array} $$

\mycaption{\label{ta:mll_TEV} Integrated cross sections for different
  cuts on the dilepton invariant mass $M_{\Pl\Pl}$ at the Tevatron.
  See caption of \refta{ta:mll_LHC} and text for details.}
\end{table}

\reftas{ta:mll_LHC}--\ref{ta:mll_TEV} 
show the analogous results for a
variation of cuts on the dilepton invariant mass
$M_{\Pl\Pl}$. The corresponding differential distributions for the
  Tevatron and the LHC at $14\TeV$ are displayed in \reffi{fi:mll_all}.
Again, all other cuts and the corresponding event selection follow our
default choice as introduced in \refse{se:cuts}.

In \reftas{ta:mll_LHC}--\ref{ta:mll_TEV}, only the cross section in
the first column includes resonant \PZ\ production. All the less
inclusive cross sections only contain the tail of the distribution
generated by (far) off-shell \PZ\ bosons and photons. Hence, the cross
sections are smaller than for comparable cuts on $p_{\rT,\mathrm{jet}}$.
Note that for a large cut on $M_{\Pl\Pl}$ the cross section is dominated
by events with a CM energy close to the $M_{\Pl\Pl}$ cut since the
leptons are mainly produced back to back and the additional jet is
relatively soft. In comparison, for a large cut on
$p_{\rT,\mathrm{jet}}$ the CM energy has to be more than twice the cut
value because the transverse momentum of the jet has to be balanced.

The dilepton invariant-mass distribution is a crucial observable at
the LHC, in particular for detector calibration and also in searches
for heavy dilepton resonances. For calibration, the $\PZ+\mathrm{jet}$
channel is the main source for boosted \PZ\ bosons close to their mass
shell resulting in high-energy leptons which are balanced by the hard
jet. Hence, the precise theoretical understanding of the $M_{\Pl\Pl}$
distribution, in particular around the \PZ-boson peak, is mandatory.

Concerning the EW corrections, the large Sudakov logarithms again
result in large negative corrections at large values of the
$M_{\Pl\Pl}$ cut. They reach the level of $-20\%$ for $M_{\Pl\Pl} >
2000\GeV$. However, they cannot be compared to earlier on-shell
computations since only off-shell \PZ\ bosons and photons contribute
in the high-$M_{\Pl\Pl}$ region as mentioned above.

\begin{sloppypar}
  As expected, the corrections for bare muons are larger since
  photons, being radiated collinearly to one of the charged leptons,
  carry away momentum and reduce the dilepton invariant mass if not
  recombined with the emitting lepton. This effect can be observed for
  any cut value in \reftas{ta:mll_LHC}--\ref{ta:mll_TEV}. However, it
  is most prominent around the \PZ-boson peak in the differential
  distribution displayed in \reffi{fi:mll_all}. For bare muons, events
  that are enhanced by muon-mass logarithms are shifted to lower bins
  in the distributions. Hence, the correction is large and negative up
  to $-20\%$ around the \PZ\ peak. The correction below the \PZ\ 
  resonance is positive. While the absolute correction compared to the
  differential cross section at the peak is relatively small, the
  relative correction to the distribution is huge because the LO
  result is relatively small away from the narrow \PZ-boson resonance.
  It amounts to more than $100\%$ around a dilepton invariant mass of
  $75\GeV$. The large correction, however, does not signal a breakdown
  of perturbation theory but only reflects the particular kinematic
  situation around the \PZ\ peak.  If photons in a small cone around a
  radiating lepton are recombined, the corrections from almost
  collinear photons outside the cone are still sizeable and amount to
  roughly half the size of the corrections for bare muons. The size of
  the corrections is further reduced if the cone size for the
  lepton--photon recombination is increased. To reach an accuracy level
  of a few percent near the resonance, multi-photon radiation
  should be included in the calculation.
\end{sloppypar}

\bfi     
\bce
\includegraphics[width=15.7cm]{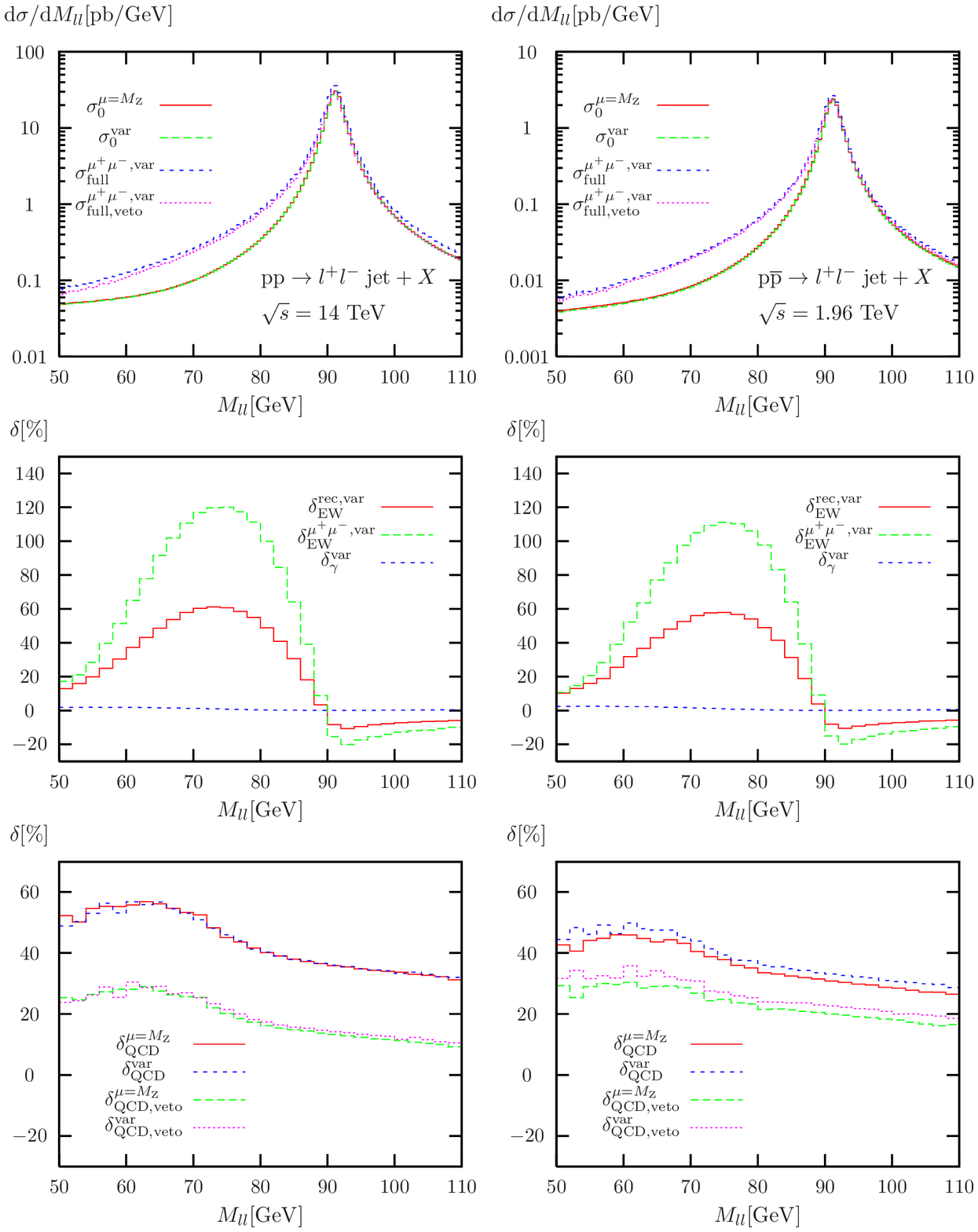} 
\ece
\mycaption{\label{fi:mll_all} LO and fully corrected distribution (top),
  corresponding relative EW and photon-induced corrections (middle), and 
  relative QCD corrections (bottom) for the 
  dilepton mass at the LHC (left) and the Tevatron (right).
  }
\efi 

\bfi     
\bce
\includegraphics[width=15.7cm]{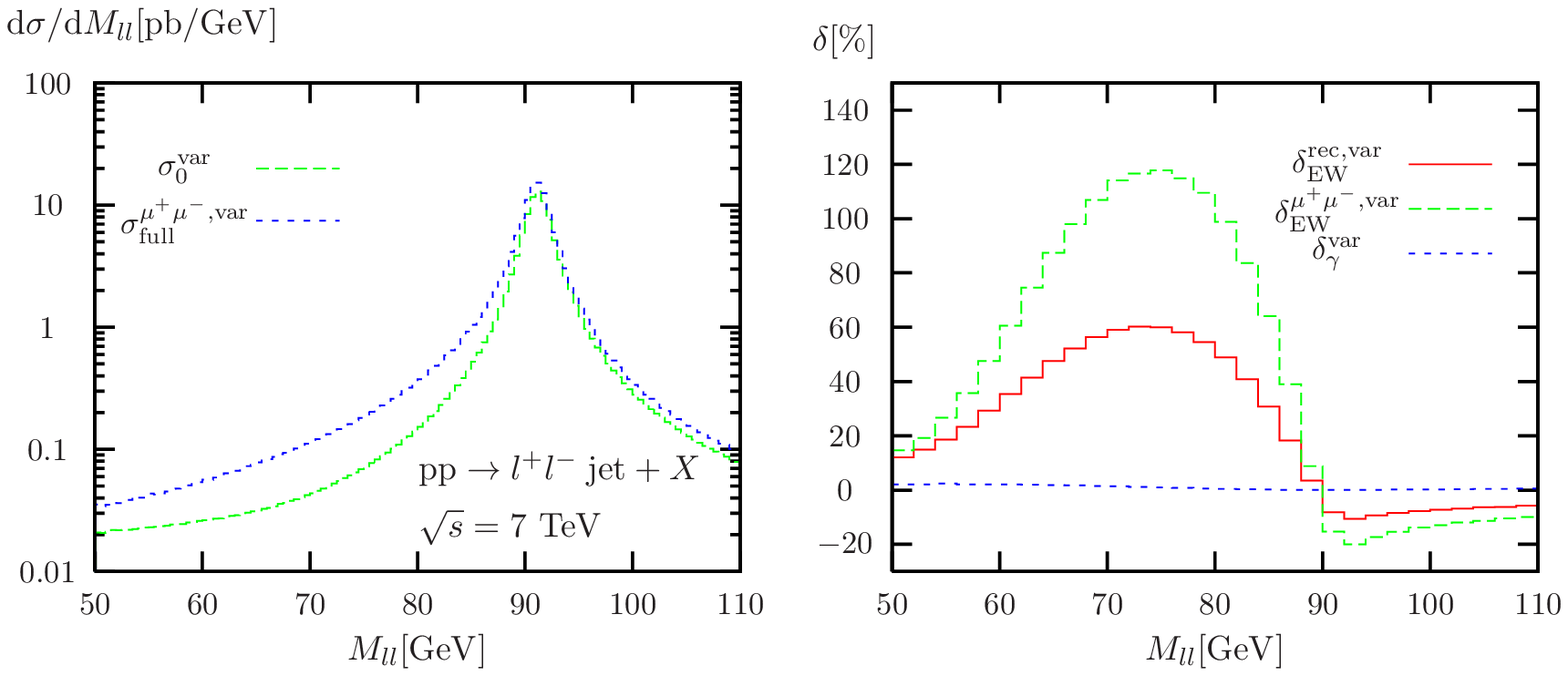} 
\ece
\mycaption{\label{fi:mll_7T} LO and fully corrected distribution (left)
  and corresponding relative EW and photon-induced corrections (right)
  for the dilepton mass at the LHC with $\sqrt{s}=7\TeV$.
  }
\efi 

The CM energy at the LHC hardly plays any role for the size of the
relative EW corrections. In \reffi{fi:mll_7T} we show the LO cross
section and EW corrections for $\sqrt{s}=7\TeV$. Only the LO cross
section strongly depends on the CM energy.  Even the results
for the EW corrections at the Tevatron (see \reffi{fi:mll_all}) hardly
differ from the ones at the LHC.

Since the dilepton mass is a property of the lepton--photon system only
and it is invariant under boosts when recoiling against a hard QCD jet,
our results can be compared to the EW corrections to the line shape of
inclusive \PZ-boson production as, for instance, recently investigated
in \citere{Dittmaier:2009cr}. It turns out that the EW corrections are
quite similar around the \PZ-boson peak. However, in the lower tail of
the distributions the corrections only reach up to $80\%$ in the
inclusive analysis (see e.g.\ Figure~12 in \citere{Dittmaier:2009cr}).  A
large part of this deviation can be attributed to the fact that the LO
result for $\PZ+\mathrm{jet}$ production drops off faster than in the
inclusive case, so that the relative corrections are different.

Turning again to NLO QCD, the corrections around the \PZ\ resonance in
the invariant-mass distribution of the leptons are pretty flat since QCD
can only indirectly affect the line shape of the colour-neutral \PZ\
boson.  Depending on the exact event definition, the corrections amount
to a few tens of percent. However, as shown in \refta{ta:mll_LHC}, in
the high-invariant-mass tail both scale choices fail to reflect the
kinematical situation, since the production of a far off-shell \PZ\
boson is dominated by the region near the threshold set by the cut on
$M_{\Pl\Pl}$.  In this region the \PZ\ boson decays mainly to
back-to-back leptons with relatively soft jet activity.  Hence,
$\delta_\mathrm{QCD}^{\mu=\MZ}$ as well as $\de_\QCD^\mathrm{var}$
become large and negative.  Consequently, a variable scale choice based
on the invariant mass of the lepton pair 
\beq \mu^{\mathrm{var},M_{ll}}
= \mu_{\mathrm{R}}^{\mathrm{var},M_{ll}} =
\mu_{\mathrm{F}}^{\mathrm{var},M_{ll}} =
\sqrt{M_{ll}^2+(p_{\mathrm{T}}^{\mathrm{had}})^2} 
\eeq 
reflects the
underlying kinematics in a better way.  Thus, for this particular
observable, we also present LO and NLO QCD predictions for this scale
choice in \reftas{ta:mll_LHC}--\ref{ta:mll_TEV}. Indeed, a significant
stabilization of the relative QCD corrections, especially for large
$M_{ll}$ cuts at the LHC, is found. The relative EW corrections do only
change insignificantly when using the alternative scale.

\subsubsection{Transverse momentum of the charged leptons}

\bfi     
\bce
\includegraphics[width=15.7cm]{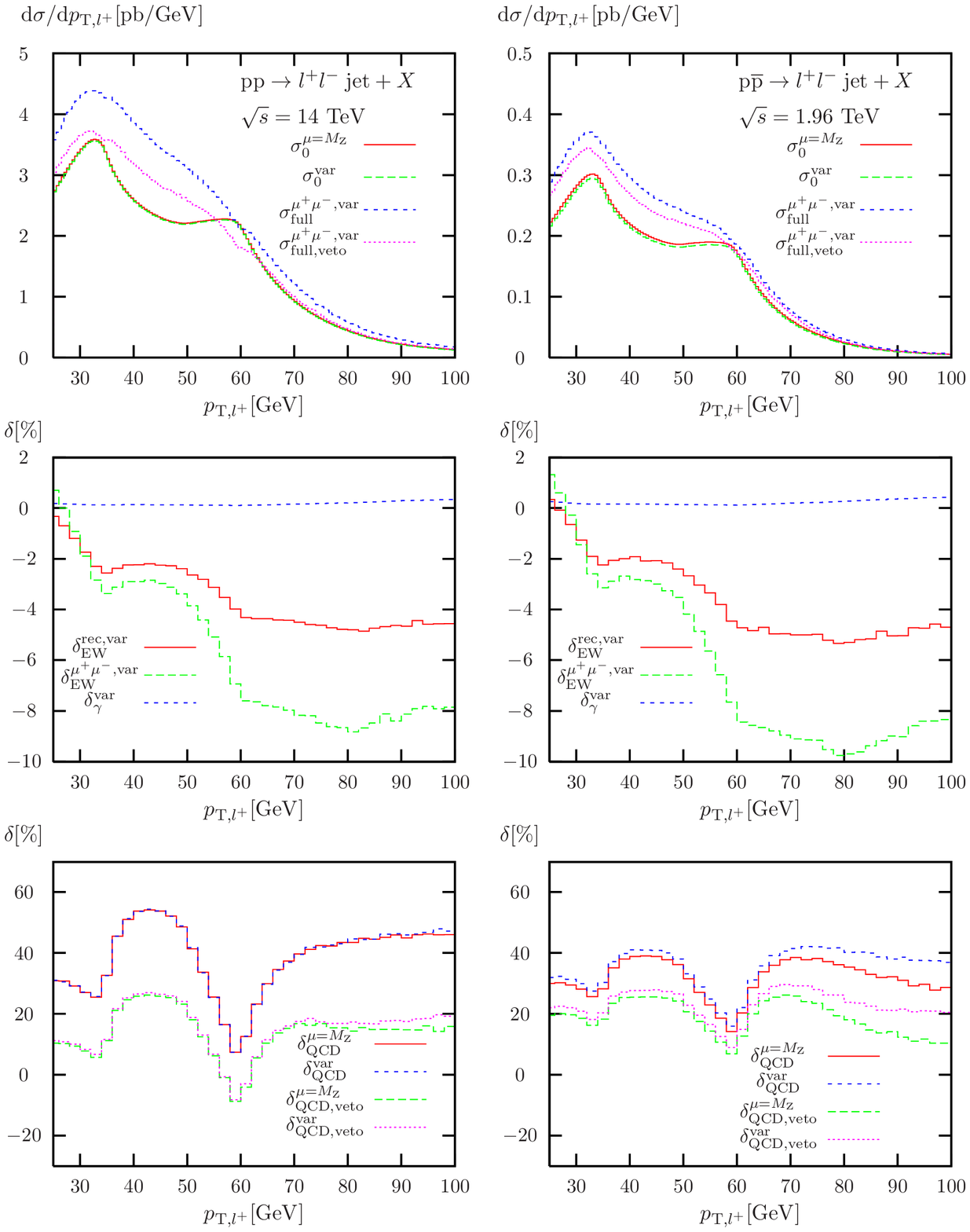} 
\ece
\mycaption{\label{fi:pt_all} LO and fully corrected distribution (top),
  corresponding relative EW and photon-induced corrections (middle), and 
  relative QCD corrections (bottom) for the transverse momentum
  of the positively charged lepton at the LHC (left) and the Tevatron (right).
  }
\efi 

In contrast to the invariant mass of the two charged leptons the
transverse momentum of each of the leptons is sensitive to the recoil
due to the hard jet in the event. Hence, the Jacobian peak of the
transverse momentum at half the \PZ-boson mass in the inclusive
\PZ-boson sample is washed out when an additional hard jet is present.
The LO results and the corresponding correction for the positively
charged lepton are shown in \reffi{fi:pt_all}. While the corrections
are equivalent for the negatively charged lepton at the Tevatron, at
the LHC they differ at the percent level but are qualitatively
similar.

The EW corrections are at the percent level and quickly increase in
size in the tail of the distribution for $p_{\rT,\Pl^+}> 60\GeV$, in
particular for bare muons where they almost reach $-10\%$. For even
higher $p_{\rT,\Pl^+}> 100\GeV$ the typical increase of the correction
due to the Sudakov logarithms can be observed. However, in contrast to
the transverse-momentum distribution of the leading jet, the
lepton--photon recombination still matters and the corrections differ
by roughly $5\%$ in the whole tail of the distribution.

The QCD corrections $\delta_\mathrm{QCD}$ for $p_{\rT,\Pl^+}$ show 
pronounced dips where the LO cross section has peaks (see
\reffi{fi:pt_all}). The real corrections do not particularly populate
the regions of the distributions that are enhanced due to the particular
LO kinematics.  For $p_{\rT,\Pl^+} < 100\GeV$,
$\delta_\mathrm{QCD}^{\mu=\MZ}$ and $\delta_\mathrm{QCD}^{\mathrm{var}}$
are practically identical.  However, for larger transverse momenta, the
corrections for the two scale choices differ significantly. Here,
$\delta_\mathrm{QCD}^{\mu=\MZ}$ grows large and negative to compensate
for the overestimated LO cross section.  This is expected, since the
hard jet recoiling against the high-$p_{\rT}$ lepton should be reflected
in the scale choice.

At the Tevatron, the kinematical features are less pronounced and the deviation
using a fixed scale starts already at $p_{\rT,\Pl^+} \gsim 70\GeV$. However, all
qualitative features for the EW as well as the QCD corrections are the
same.

\subsubsection{Transverse mass of the charged leptons}

For \PZ-boson production, the transverse mass does not play a central
role because the lepton system can be fully reconstructed. However, it
is instructive to compare the transverse-mass distribution for
$\PZ+\mathrm{jet}$ production with the corresponding distribution for
$\PW+\mathrm{jet}$.  For $\PW+\mathrm{jet}$, due to the neutrino in the
final state, only the transverse mass but not the invariant mass of the
two leptons can be measured. The distribution is displayed in
\reffi{fi:mt_all}. It shows all the features of the analysis for
$\PW+\mathrm{jet}$ discussed in detail in
\citere{Denner:2009gj}. However, the EW corrections are roughly a factor
of two larger and reach the level of $-20\%$ for bare muons in the
interesting region around the Jacobian peak, simply because there are
two charged leptons in the final state which radiate photons. This 
  example indicates that one has to be especially careful when
using ratios of \PW\ and \PZ\ cross sections with the aim to reduce 
theoretical and parametric (such as PDF) uncertainties.
Even though the
two systems are almost equivalent from the point of view of QCD, EW
corrections differ significantly.

\bfi     
\bce
\includegraphics[width=15.7cm]{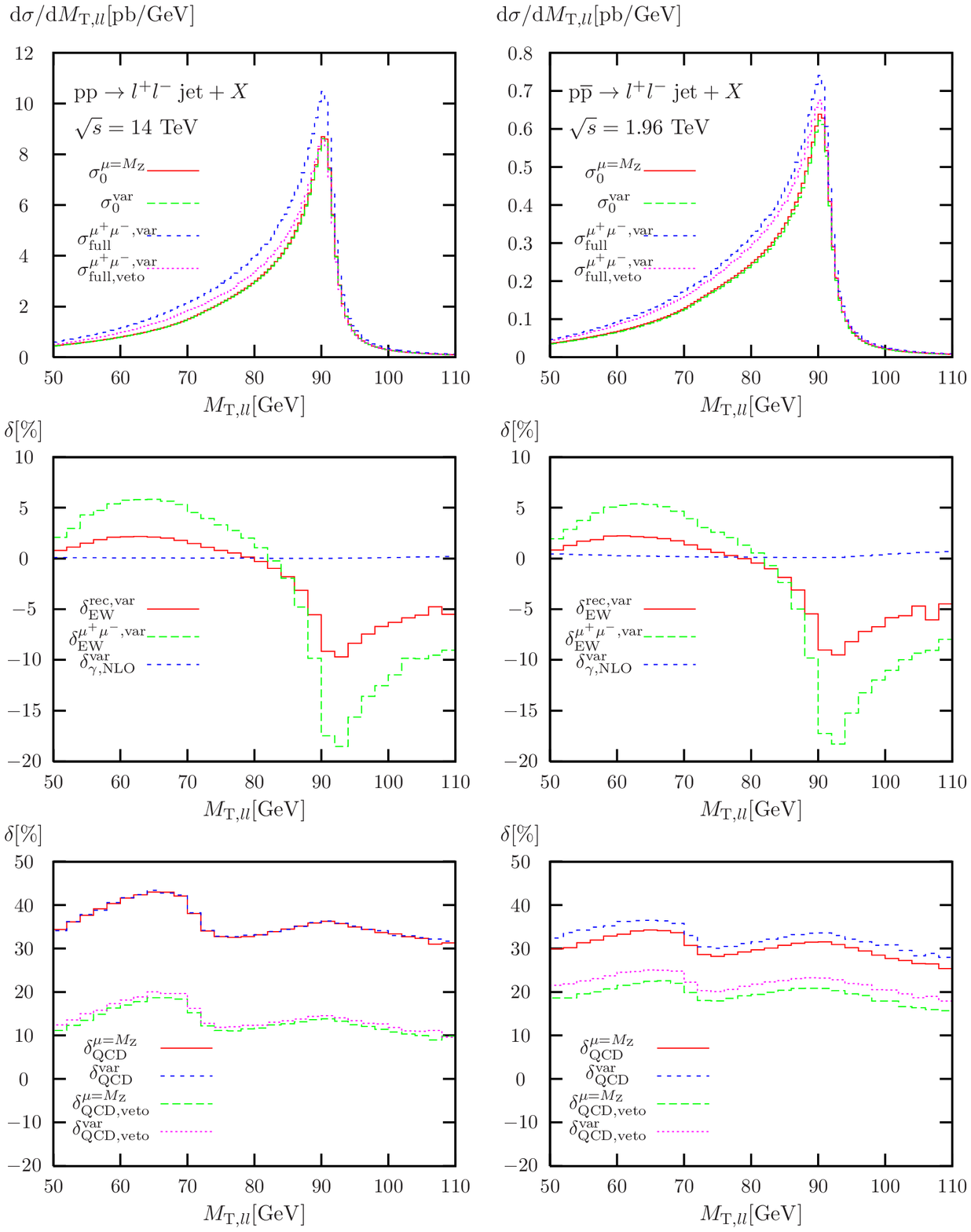} 
\ece
\mycaption{\label{fi:mt_all} LO and fully corrected distribution (top),
  corresponding relative EW and photon-induced corrections (middle), and 
  relative QCD corrections (bottom) for the transverse mass
  of the two charged leptons at the LHC (left) and the Tevatron (right).
  }
\efi 

\subsubsection{Rapidity distributions}

In \reffi{fi:yl_all}, we display the results on the rapidity
distribution for the positively charged lepton. For the LHC, the EW
and QCD corrections are rather flat and only slightly increase the
differential cross section in the forward and backward regions. The
distribution and the corrections are symmetric with respect to
$y_{l^+}=0$.  At the Tevatron, this symmetry is only slightly disturbed
at LO and NLO QCD.  However, the EW corrections are asymmetric, large
and negative for $y_{l^+}=-2.5$, and almost zero for $y_{l^+}=2.5$. Of
course, the distribution for the rapidity of the negatively charged
lepton shows the analogous behaviour with reversed rapidities w.r.t.\
the case of the $l^+$.

Concerning the rapidity of the leading jet $y_{\mathrm{jet}}$, the EW
corrections are flat and do not disturb the LO shapes of the
distribution, as can be seen in \reffi{fi:yj_all}. The QCD corrections
are only slightly  larger in the forward and backward regions
compared to the central region at the LHC.

Since both leptons can be fully reconstructed, also the rapidity
$y_{\Pl\Pl}$ of the final-state lepton pair---which resembles the
rapidity of the
intermediate boson---is experimentally accessible. The corresponding
results are shown in \reffi{fi:yZ_all}.
\bfi     
\bce
\includegraphics[width=15.7cm]{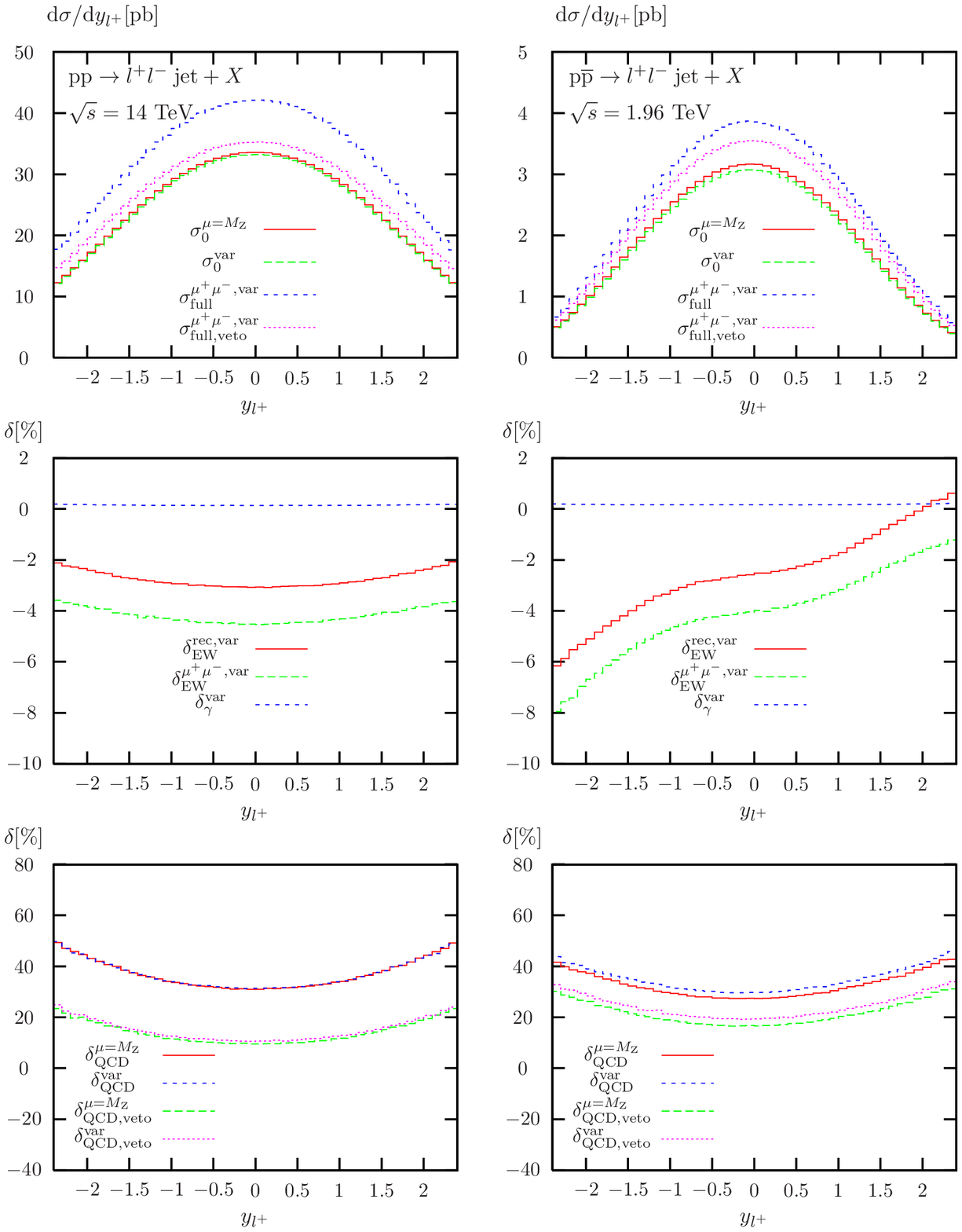} 
\ece
\mycaption{\label{fi:yl_all} LO and fully corrected distribution (top),
  corresponding relative EW and photon-induced corrections (middle), and 
  relative QCD corrections (bottom) for the rapidity
  of the positively charged lepton at the LHC (left) and the Tevatron (right).
  }
\efi 

\bfi     
\bce
\includegraphics[width=15.7cm]{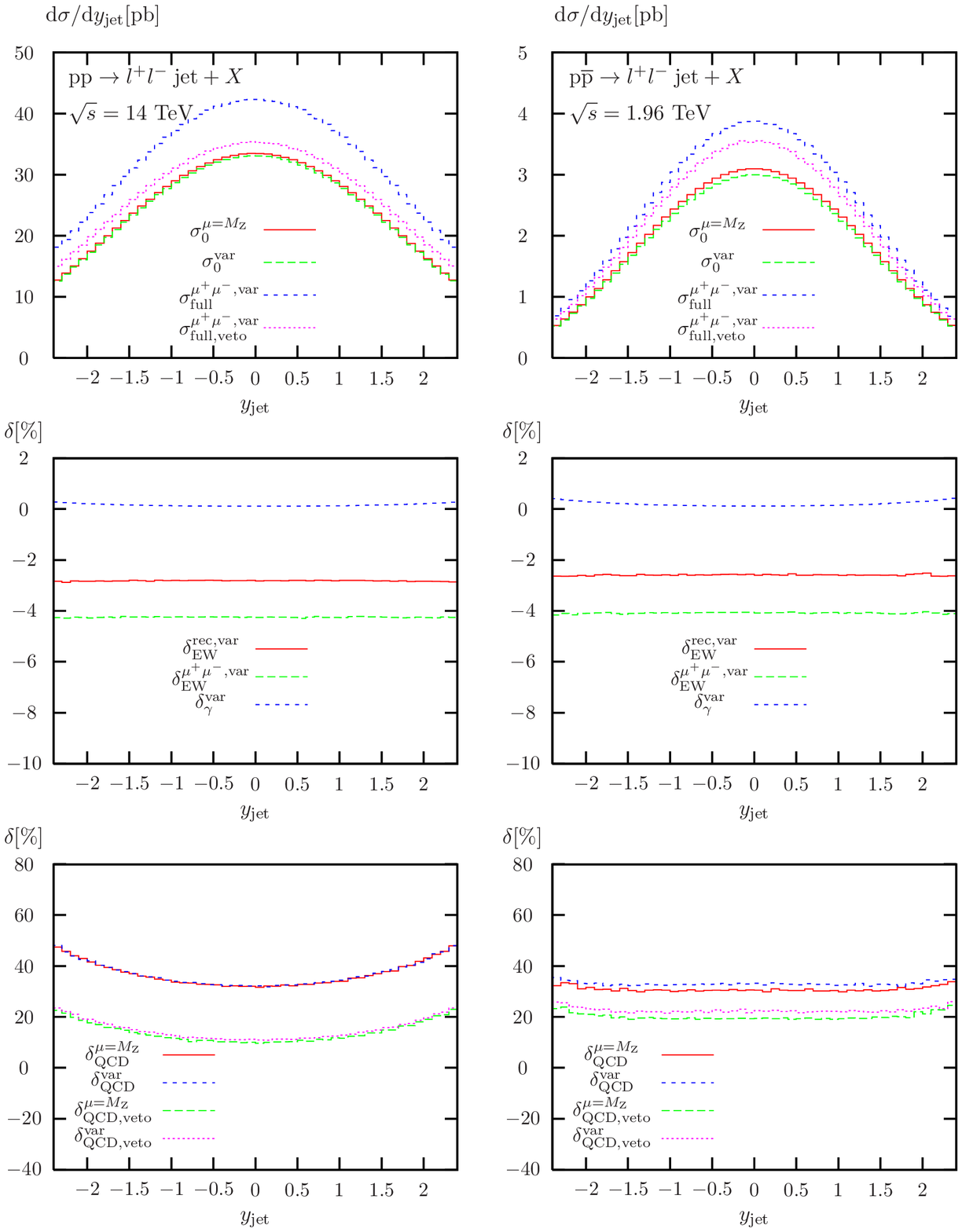} 
\ece
\mycaption{\label{fi:yj_all} LO and fully corrected distribution (top),
  corresponding relative EW and photon-induced corrections (middle), and 
  relative QCD corrections (bottom) for the rapidity
  of the leading jet at the LHC (left) and the Tevatron (right).
  }
\efi 

\bfi     
\bce
\includegraphics[width=15.7cm]{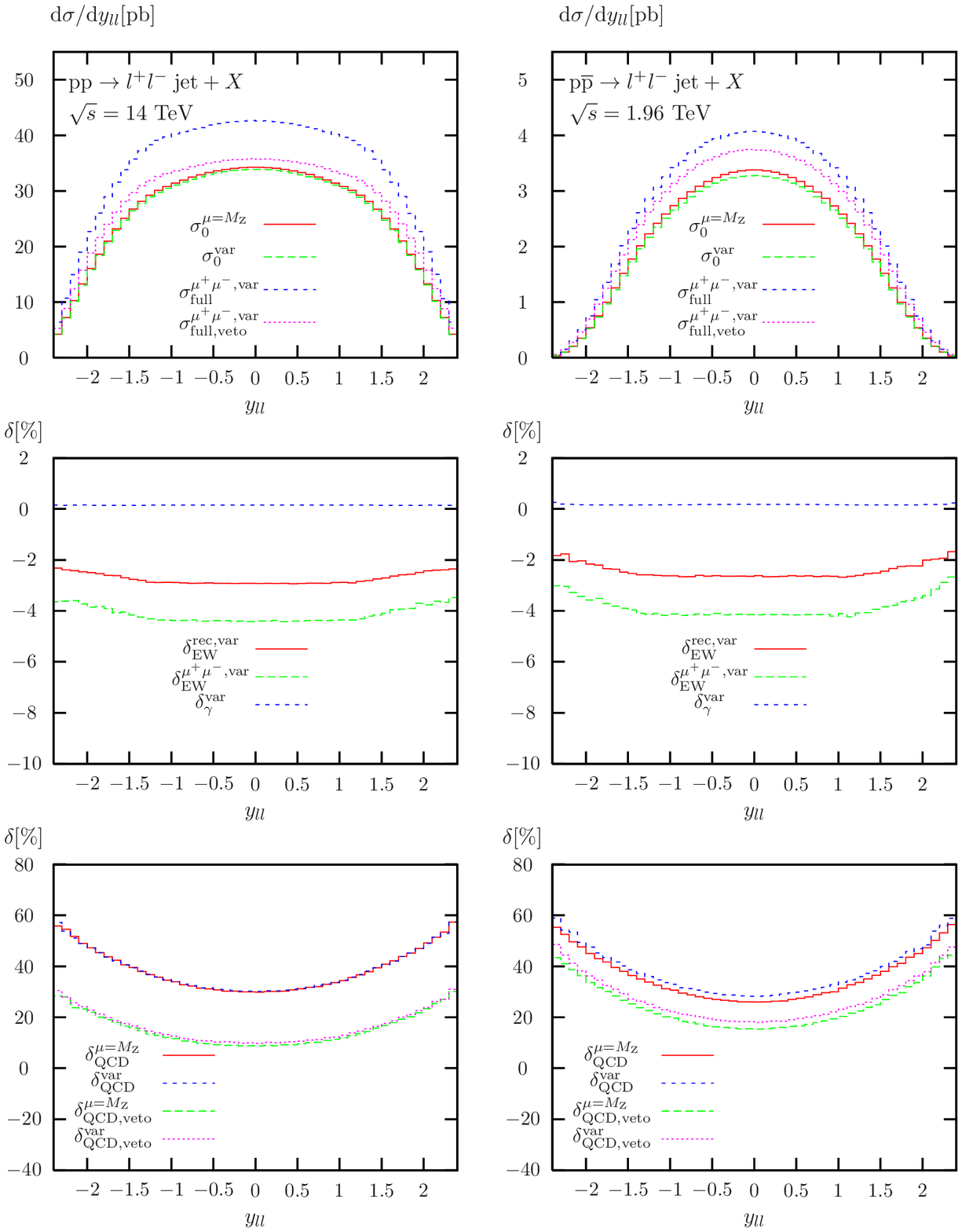} 
\ece
\mycaption{\label{fi:yZ_all} LO and fully corrected distribution (top),
  corresponding relative EW and photon-induced corrections (middle), and 
  relative QCD corrections (bottom) for the rapidity
  of the dilepton system at the LHC (left) and the Tevatron (right).
  }
\efi 

\section{Conclusions}
\label{se:concl}

Following our study on $\PW+\mathrm{jet}$ production~\cite{Denner:2009gj},  
we have presented the first calculation of the full EW NLO
corrections to the production of two opposite-sign charged leptons 
in association with a hard jet at hadron colliders. For many observables
the cross section is dominated by on-shell $\PZ+\mathrm{jet}$ production with 
a subsequent leptonic \PZ-boson decay. However, in our calculation
all off-shell effects as well as the contributions of and the
interference with an intermediate photon are taken into account, so
that also observables that are not dominated by on-shell \PZ\ bosons
are described with NLO accuracy in view of both QCD and EW
corrections.

We have implemented our results in a flexible Monte Carlo code which
can model the experimental event definition at the NLO parton level.
The distinction of $\PZ+\mathrm{jet}$ and $\PZ+\mathrm{photon}$
production is consistently implemented by making use of the measured
quark-to-photon fragmentation function.  We have also recalculated the
NLO QCD corrections supporting a phase-space-dependent scale choice.
Photon-induced processes are included at leading order but turn out to
be phenomenologically unimportant.

The presented EW corrections are particularly large for the \PZ-boson
line shape, i.e.\ the dilepton-invariant-mass distribution, mainly due
to collinear final-state radiation. At the peak of the distribution,
the corrections reach $-20\%$ for bare muons while they are at the
order of $100\%$ in the lower tail of the distribution, even larger
than for the inclusive \PZ-boson line shape where no additional hard
jet in the final state is demanded. When the EW corrections are not
enhanced by final-state radiation due to particular kinematics, they
are typically negative and at the level of a few percent. However, in
the tail of distributions which are sensitive to large CM energies,
the well-known Sudakov logarithms become dominant, and the EW
corrections increase up to $-25\%$ at partonic $\sqrt{s}\sim 2\TeV$.
For the $p_\rT$
distribution of the jet, these results agree with earlier results in
the on-shell approximation for the \PZ\ boson~\cite{Kuhn:2005az}. The
QCD corrections have a typical size of a few tens of percent. However,
they can become extremely large (hundreds of percent) at large jet
$p_\rT$ unless a sensible veto on a second hard jet is applied.
The presented integrated cross sections and differential distributions
demonstrate the applicability and flexibility of our setup which will
be useful for accurate predictions of any observable in the
investigated final state.

The importance of the neutral-current Drell--Yan process as a testing
ground for perturbative calculations and for understanding and
calibrating the detectors at the LHC can hardly be overestimated. Our
calculation extends the availability of theoretical predictions for
this process class to the EW corrections to associated production of
\PZ\ bosons with a hard jet.  Our calculation also constitutes one
important part of a full NNLO prediction of the mixed EW and QCD
corrections for inclusive \PZ\ production. In the range of
intermediate and large transverse momenta of the additional hard jet
our calculation delivers state-of-the-art predictions, for small
transverse momenta the pure NLO calculation should of course be
improved by dedicated QCD resummations, a task that goes beyond the
scope of this paper.

\subsection*{Acknowledgements}

This work is supported in part by the Gottfried Wilhelm Leibniz programme of
the Deutsche Forschungsgemeinschaft (DFG), by the 
DFG Sonderforschungsbereich/Transregio 9 ``Computergest\"utzte 
Theoretische Teilchenphysik'', and by
the European Community's Marie-Curie Research Training Network
HEPTOOLS under contract MRTN-CT-2006-035505.

\end{document}